\documentclass[universe,review,accept,pdftex,oneauthor]{mdpi}
\firstpage{1}
\pubvolume{8}
\issuenum{9}
\articlenumber{480}
\pubyear{2022}
\copyrightyear{2022}
\externaleditor{Academic Editors: Robert H. Bernstein and Bertrand Echenard
} 
\datereceived{11 April 2022}
\dateaccepted{18 August 2022}
\datepublished{13 September 2022}
\hreflink{https://doi.org/10.3390/\linebreak universe8090480} 
\pdfoutput=1

\usepackage{threeparttable} 
\usepackage{longtable}
\usepackage{array}
\usepackage{xspace}
\usepackage[normalem]{ulem}	
\Title{Searches for Lepton Flavor Violation in Tau Decays at Belle II}

\TitleCitation{Searches for Lepton Flavor Violation in Tau Decays at Belle II}

\Author{Swagato Banerjee \href{https://orcid.org/0000-0001-8852-2409}{\orcidicon}}

\AuthorNames{Swagato Banerjee}

\AuthorCitation{Banerjee, Sw.} 

\address[1]{Department of Physics and Astronomy, University of Louisville, Louisville, KY 40292, USA; swagato.banerjee@louisville.edu}

\abstract{Searches for lepton flavor violation in tau decays are unambiguous signatures of new physics.
The branching ratios of tau leptons at the level of $10^{-10}$--$10^{-9}$ can be probed using 50~$ab^{-1}$
of electron-positron annihilation data being collected by the Belle II experiment at the world’s highest luminosity accelerator, the SuperKEKB,
located at the High Energy Accelerator Research Organization, KEK, in Tsukuba, Japan.
Searches with such expected sensitivity will either discover new physics or strongly constrain several new physics models.
}


\usepackage{relsize}
\def\babar{\mbox{\slshape B\kern-0.1em{\smaller A}\kern-0.1em
B\kern-0.1em{\smaller A\kern-0.2em R}}}
\def\eff{\epsilon}

\def\GeV{\ifmmode{\mathrm{Ge\kern -0.1em V}}\else
\textrm{Ge\kern -0.1em V}\fi}
\def\MeV{\ifmmode{\mathrm{Me\kern -0.1em V}}\else
\textrm{Me\kern -0.1em V}\fi}

\newcommand{\mmm}     {\ensuremath{\mu^-\!\mu^+\!\mu^-}}

\def\mec        {\mathrm{m}_{\mathrm EC}}
\def\DeltaE     {\Delta {\rm{E}}}

\def\taumg      {\tau^- \to \mu^- \gamma}
\def\taueg      {\tau^- \to e^- \gamma}
\def\taulg      {\tau^- \to \ell^- \gamma}

\def\taulll     {\tau^- \to \ell^- \ell^+ \ell^-}

\def\BR         {{\cal{B}}}

\def\BRtaulg    {\BR(\taulg)}
\def\BRtaulll   {\BR(\taulll)}

\def\mmm       {\mu^- \mu^+ \mu^-}

\def\taummm    {\tau^- \to \mmm}

\def\invfb   {\ensuremath{\mbox{\,fb}^{-1}}\xspace}
\def\invab   {\ensuremath{\mbox{\,ab}^{-1}}\xspace}

\newcommand{\gev}{\ensuremath{\mathrm{\,Ge\kern -0.1em V}}\xspace}
\newcommand{\mev}{\ensuremath{\mathrm{\,Me\kern -0.1em V}}\xspace}

\def\mec        {\mathrm{m}_{\mathrm EC}}
\def\DeltaE     {\Delta {\rm{E}}}

\def\GeV{\ifmmode{\mathrm{Ge\kern -0.1em V}}\else
\textrm{Ge\kern -0.1em V}\fi}
\def\MeV{\ifmmode{\mathrm{Me\kern -0.1em V}}\else
\textrm{Me\kern -0.1em V}\fi}

\begin{document}


\section{Introduction}

Lepton flavor conservation stands out in the Standard Model (SM)
among all other symmetries
because it is not associated with any underlying conserved current.
Lepton flavor violation (LFV) in the charged sector is predicted
by many new physics (NP) models.
The small but finite mass of the neutrinos in the SM allow charged LFV
in neutrino-less two-body decays,
e.g., $\tau^- \to \ell_i^- \gamma$ decays~(charge conjugate modes are implied throughout the text, unless otherwise specified),
where $\ell_i^-$ ($i$ = 1, 2) denotes light charged leptons ($e^-$, $\mu^-$).
However, such decays are suppressed
by a factor of $({m_\nu^2}/{m_W^2})^2$~\cite{Lee:1977tib},
which produces experimentally unreachable rates of the order of $10^{-54}$.
For neutrino-less three-body decays,
e.g., $\tau^- \to \ell_i^- \ell_j^+ \ell_a^- $ decays
(where $a~=~i~\mathrm{or}~j$, and~$i$ may or may not be equal to $j$),
two conflicting predictions existed in the literature:
one of the order of $10^{-55}$~\cite{Petcov:1976ff}
and another of the order of $10^{-14}$~\cite{Pham:1998fq}.
Recently, the~contributions due to finite neutrino masses for such decays
were re-scrutinized and found to be in the range of $[10^{-56},10^{-54}]$~\cite{Hernandez-Tome:2018fbq,Blackstone:2019njl},
thereby laying to rest the claim that such decays could be
of the order of $10^{-14}$ in the SM.
Thus, any observation of charged LFV is an unambiguous signature of~NP.

{{{A discovery}}} of LFV in the charged sector
may {{{provide deeper insight into}}} several unsolved mysteries
such as the origin of the dark sector, large baryon number asymmetry,
number of flavor generations and extra dimensions.
Many NP models, such as low-scale seesaw models~\cite{Cvetic:2002jy},
supersymmetric standard models~\cite{Ellis:1999uq, Ellis:2002fe, Dedes:2002rh, Brignole:2003iv, Masiero:2002jn, Hisano:2009ae,Fukuyama:2003hn},
little Higgs models~\cite{Choudhury:2006sq, Blanke:2007db},
leptoquark models~\cite{Davidson:1993qk},
non-universal $Z^\prime$ models~\cite{Yue:2002ja},
and extended Higgs models~\mbox{\cite{Akeroyd:2009nu, Harnik:2012pb, Celis:2013xja, Omura:2015nja, Goudelis:2011un}},
predict LFV in $\tau$ decays at the level of $10^{-10}$--$10^{-8}$
which are just below the current experimental bounds.
Predictions for two-body and three-body neutrino-less $\tau$ decays
from some of these
NP models~\cite{Cvetic:2002jy,Ellis:1999uq,Ellis:2002fe,Dedes:2002rh,Brignole:2003iv,Masiero:2002jn,Fukuyama:2003hn,Yue:2002ja}
are tabulated in Table~\ref{taulfv_predictions}.

\begin{table}[H]
\setlength{\tabcolsep}{8.15mm}

\caption{The branching fractions ($\BR$) for $\taulg$ and $\taulll$ decays in some NP~models.}
\label{taulfv_predictions}
\begin{tabular}{lcc}
\noalign{\hrule height 1pt}

& {\boldmath{$\BRtaulg$}} & {\boldmath{$\BRtaulll$}}\\\midrule
SM + seesaw~\cite{Cvetic:2002jy}                    & $10^{-9}$    & $10^{-10}$ \\
SUSY + Higgs~\cite{Dedes:2002rh,Brignole:2003iv}    & $10^{-10}$   & $10^{-8}$  \\
SUSY + SO(10)~\cite{Masiero:2002jn,Fukuyama:2003hn} & $10^{-8}$    & $10^{-10}$ \\
Non-universal Z$^\prime$~\cite{Yue:2002ja}          & $10^{-9}$    & $10^{-8}$  \\ \noalign{\hrule height 1pt}
\end{tabular}
\end{table}

Since $\tau$ is the only lepton that can decay hadronically,
many neutrino-less final states are allowed via LFV processes in NP models
within the observable parameter space~\cite{Black:2002wh}.
Example of such processes include final states containing
a light lepton and a meson
e.g., $\tau^- \to \ell^- M^0$ (where $M^0 = \pi^0, K_S^0, \eta, \rho^0, \omega,
K^{\star0}, \bar{K}^{\star0}, \eta^\prime, a_0, f_0, \phi$),
or a light lepton and two mesons: $\tau^- \to \ell^- h^+h^-, \;
\ell^+ h^- h^-$, \; $\ell^- h^0 h^0$
(where $h^\pm = \pi^\pm/K^\pm, h^0 =  \pi^0/K_S^0$).

Searches for all possible LFV processes in decays of $\tau$ are necessary
because there are strong correlations
between the expected rates of the different channels in various models.
For example, in~some supersymmetric seesaw models~\cite{Babu:2002et, Sher:2002ew},
the relative rates of
${\cal{B}}(\tau^\pm \to \mu^\pm \gamma)$ :
${\cal{B}}(\tau^\pm \to \mu^\pm \mu^+\mu^-)$ :
${\cal{B}}(\tau^\pm \to \mu^\pm \eta)$
are predicted to have specific ratios,
depending on the model parameters.
In the unconstrained minimal supersymmetric model,
which includes various correlations between the $\tau$ and $\mu$ LFV rates,
the LFV branching fractions of the $\tau$ lepton in some decay channels
can be as high as $10^{-8}$~\cite{Brignole:2004ah, Goto:2007ee}),
while still respecting the strong experimental bounds
on the LFV decays of the $\mu$ lepton~\cite{MEG:2016leq,SINDRUM:1987nra}.
Thus, it is critical to probe all possible LFV modes of the $\tau$ lepton,
because any excess in a single channel will not provide sufficient information
to nail down the underlying LFV mechanism or even to identify an underlying~theory.

More exotic decay modes, such as $\tau^-\to\pi^+\ell^-\ell^-$ and $K^+\ell^-\ell^-$,
accompanied by a violation of the lepton number (LNV),
are predicted at the level of $10^{-10}$--$10^{-8}$ in several scenarios
beyond the SM~\cite{LopezCastro:2012udb}.
Several of these decay modes are expected to have branching ratios
close to existing experimental limits in NP models,
e.g., heavy Dirac \mbox{neutrinos~\cite{Gonzalez-Garcia:1991brm, Ilakovac:1999md}},
supersymmetric processes~\cite{Sher:2002ew, Arganda:2008jj},
flavor-changing $Z^\prime$ exchanges with non-universal couplings~\cite{Li:2005rr},
etc., to name a few.
Wrong-sign $\tau^-\to\ell_i^+\ell_j^-\ell_j^-$ decays are very intriguing
because they are expected at rates only one order of magnitude below the present bounds
in some NP models, e.g.,~the Littlest Higgs model
with T-parity realizing an inverse seesaw~\cite{Pacheco:2021djh}.

Most models for baryogenesis, a~hypothetical physical process
based on different descriptions of the interaction between the fundamental particles
that took place during the early universe
producing the observed matter--antimatter asymmetry,
require baryon number violation (BNV),
which in charged lepton decays automatically implies LNV and LFV{{~\cite{LHCb:2013fsr}}}.
Angular momentum conservation requires the difference of
net baryon number (B) and lepton number (L) to be equal to either 0 or 2.
Although the SM conserves this difference,
the symmetry group for the sum of baryon number and lepton number
can be associated with an anomalous current.
A set of models predicts baryogenesis that conserves B-L
but includes instanton induced B+L violating currents~\cite{Fuentes-Martin:2014fxa}.
In a large class of models~\cite{Hou:2005iu},
BNV in $\tau$ decay modes containing baryons in the final state, for~example,
$\tau^- \to \pi^- \Lambda, \pi^- \bar{\Lambda}, K^- \Lambda, K^- \bar{\Lambda},
\bar{p} \gamma, \bar{p} \ell_i \bar{\ell}_j ~\mathrm{and}~ p \ell_i \ell_j$,
are predicted at observable rates in the large $\tau$ data set
that the Belle II detector will record over the coming~years.


\section{Belle II Experiment at~SuperKEKB} 

The most restrictive limits on LFV in $\tau$ decays at the level of $10^{-8}$ have been obtained by the first generation of the B-Factory experiments, Belle and \babar,
where a big data sample of $\tau$' s was generated
thanks to large and similar values of the production cross-sections of
$B-$ mesons and $\tau-$ pairs around the $\Upsilon(4S)$ resonance
at the level of a nanobarn ($nb$)~\cite{Banerjee:2007is}.
Belle and \babar\ experiments collected approximately
one attobarn-inverse ($ab^{-1}$) and half an $ab^{-1}$ of $e^-e^+$ annihilation data,
respectively.
The next generation of the B-Factory experiment, Belle II,
is expected to collect 50 $ab^{-1}$ of data over the next decade~\cite{Belle-II:2022cgf}.
Such a huge data sample corresponding to $10^{11}$ single $\tau$-decays
would lower the limits on LFV in $\tau$ decays by one or two orders of~magnitude.

\subsection{Luminosity Upgrade of~SuperKEKB} 

The asymmetric beam energy $e^-e^+$ collider, SuperKEKB,
is an upgrade of the KEKB accelerator facility in Tsukuba, Japan,
and has a circumference of about 3~km.
The main components of the SuperKEKB collider complex
are a 7~GeV electron ring known as the high-energy ring (HER),
a 4~GeV positron ring known as the low-energy ring (LER),
and an injector linear accelerator with a 1.1~GeV positron damping ring~\cite{Akai:2018mbz}.
The HER and the LER have four straight sections named Tsukuba, Oho, Fuji, and~Nikko,
with the interaction point in the straight section of Tsukuba,
where the Belle II detector is~located.

The target integrated luminosity of 50 $ab^{-1}$
to be collected by the Belle II experiment
will be achieved by increasing the instantaneous luminosity by a factor of 30.
Two major upgrades account for this increase:
a modest two-fold increase in the beam currents,
and a fifteen-fold reduction of the vertical beta function
at the interaction point $(\beta^\star_y)$ from 5.9~mm to 0.4~mm,
according to the “nano-beam” scheme described~below.

Compared to KEKB, the~asymmetry between the beam energies for the HER/LER beams
were reduced from 8.0/3.5~GeV to 7.0/4.0~GeV,
which reduces the beam loss due to Touschek scattering.
This also improves the solid-angle acceptance of the experiment,
which helps to analyze events with large missing energy.
Additionally, the~effects of synchrotron radiation as a result of higher currents
are mitigated. Since synchrotron radiation is proportional to the product of beam current
and the fourth power of beam energy,
the HER at SuperKEKB emits $(7/8)^4 = 59\%$
as much synchrotron radiation per unit of beam current
compared to KEKB~\cite{Wu:2015gta}.
This facilitates the SuperKEKB collider to operate at a beam current
twice the value of the~KEKB.

The very high luminosity environment of SuperKEKB required
significant upgrades of the injection beams with high current and low emittance.
The upgraded accelerator complex houses a new electron-injection gun
and a new target for positron production.
A new damping ring was installed for injection of the positron beam
with low-emittance, as~well as for improving simultaneous top-up injections
needed for the high luminosity upgrade.
The upgrade also features completely redesigned lattices for the LER and HER,
replacement of short dipoles with longer ones in the LER,
a new Titanium Nitride coated beam pipe with antechambers
to suppress the electron-cloud effect, a~modified RF system,
and a completely redesigned interaction region~\cite{Suetsugu:2016twp}.

The design of the beam parameters at SuperKEKB~\cite{Ohnishi:2013fma}
follows the “nano-beam” and the “crab-waist” schemes,
which were originally proposed for the SuperB-Factory in Italy~\cite{SuperB:2007lel}.
Accordingly, the~transverse sizes of the beam bunches
in the horizontal plane ($\sigma_x$) are squeezed to have very small values
and made to collide at a larger horizontal crossing angle
($2\phi_x$) $=$ $83~{\rm{mrad}}$ at Belle II, instead of =22~{\rm{mrad}} at Belle.
Thus, the~effective size of the overlap region ($\tilde{\sigma}_z$)
is much shorter than  what it would have been in the case of a normal head-on collision,
which is given by the longitudinal size of the beam bunches
in the horizontal plane ($\sigma_z$)~\cite{Ohnishi:2021ei}.

The vertical beta function at the interaction point (IP) is constrained
due to the hour-glass effect as:
$\beta^\star_y > \tilde{\sigma}_z = \frac{\sigma^\star_x}{\sin \phi_x} =
\frac{\sigma_z}{\Phi}$, where $\Phi$ is the large Piwinski angle.
With $\sigma_z$ of the order of 6~mm, the~$\beta^\star_y$ is thus squeezed down
to about $400~{\upmu}{\rm{m}}$, which is much shorter than the real bunch length.
In addition to increasing the luminosity,
a reduction of the interaction region of the colliding beams
restricts the vertex position along the beam axis,
thus providing an additional benefit
of more precise estimation of the primary vertex,
which helps in the reconstruction of the complete event topology
during physics~analysis.

The instantaneous luminosity in an $e^-e^+$ collider is
$${\cal{L}} = \frac{N_{e^{-}} N_{e^{+}} f}{4\pi \sigma^\star_x \sigma^\star_y} R_{\cal{L}},$$
where $N_{e^{-}}$ is the number of electrons per bunch,
$N_{e^{+}}$  is the number of positrons per bunch,
$f$ is the collision frequency of the bunch,
$\sigma^\star_x$ and $\sigma^\star_y$ are the transverse beam-profile sizes at the IP,
and $R_{\cal{L}}$ is the luminosity-reduction factor (of the order of unity)
due to the finite beam-crossing angle.
In terms of beam currents $I_{\pm} = N_{e^{\pm}} e f$, the~luminosity becomes
$${\cal{L}} = \frac{N_{e^{\mp}}I_{e^{\pm}}}{e} \frac{1}{4\pi} \frac{ R_{\cal{L}}}{\sigma^\star_x \sigma^\star_y},$$
where $e$ is the charge of the electron. Each beam affects the stability of the other,
which can be characterized by the beam--beam tune shift parameters given by
$$\xi_{(x,y)}^{\pm} =
\frac{r_e}{2\pi\gamma_{\pm}}
\frac{N_{e^{\mp}} \beta^\star_{(x,y)}}{\sigma^{\star}_{(x,y)} \big(\sigma^{\star}_{x} + \sigma^{\star}_{y}\big)}
R_{{\xi}_{(x,y)}},$$
where $r_e$ is the classical radius of the electron, $\gamma_{\pm}$ is the relativistic gamma factor of $e^{-}(e^{+})$ beams,
and $R_{{\xi}_{(x,y)}}$ is the geometric reduction factor (also of the order of unity) due to the hour-glass effect.
Putting all these factors altogether,
we arrive at the following expression for  instantaneous luminosity:
$${\cal{L}} = \frac{\gamma_{\pm}}{2er_e}
\Big( 1 + \frac{\sigma^\star_y}{\sigma^\star_x}\Big)
\frac{I_{e^{\pm}}\xi_{y}^{\pm}}{\beta^\star_y}
\Big( \frac{R_{\cal{L}}}{R_{{\xi}_{y}^\pm}} \Big),$$
where the design parameters for beam--beam tune shifts are
$\xi_{(x,y)}$ $=$ $(0.0012,~0.0807)$ for HER and $=$ $(0.0028,~0.0881)$ for LER~\cite{Ohnishi:2013fma}.
The horizontal/vertical beam sizes at the IP are reduced from
$\sigma_x^\star/\sigma_y^\star$ = 170~$\upmu$m/940~{nm} for HER  and
=147~$\upmu$m/940~{nm} for LER at Belle to
=10.7~$\upmu$m/62~{nm} for HER  and
=10.1~$\upmu$m/48~{nm} for LER at Belle II, respectively~\cite{Akai:2018mbz}.
The beam currents for Belle were 1.19~{\rm{A}}
and 1.64~{\rm{A}} for HER and LER, respectively,
compared to the design values of 2.6~{\rm{A}}
and 3.6~{\rm{A}} for HER and LER, respectively,
at Belle II~\cite{Akai:2018mbz}.
This allows one to improve upon the value of instantaneous luminosity
from \linebreak ${\cal{L}} = 2.1 \times 10^{34}{~\rm{{cm^2}{s^{-1}}}}$ at Belle to
$6.5 \times 10^{35}{~\rm{{cm^2}{s^{-1}}}}$ at Belle II~\cite{Belle-II:2022cgf}.

\subsection{Detector Upgrade of Belle~II}

\begin{figure}[H]
\centering
\includegraphics*[width=\textwidth,height=.3\textheight]{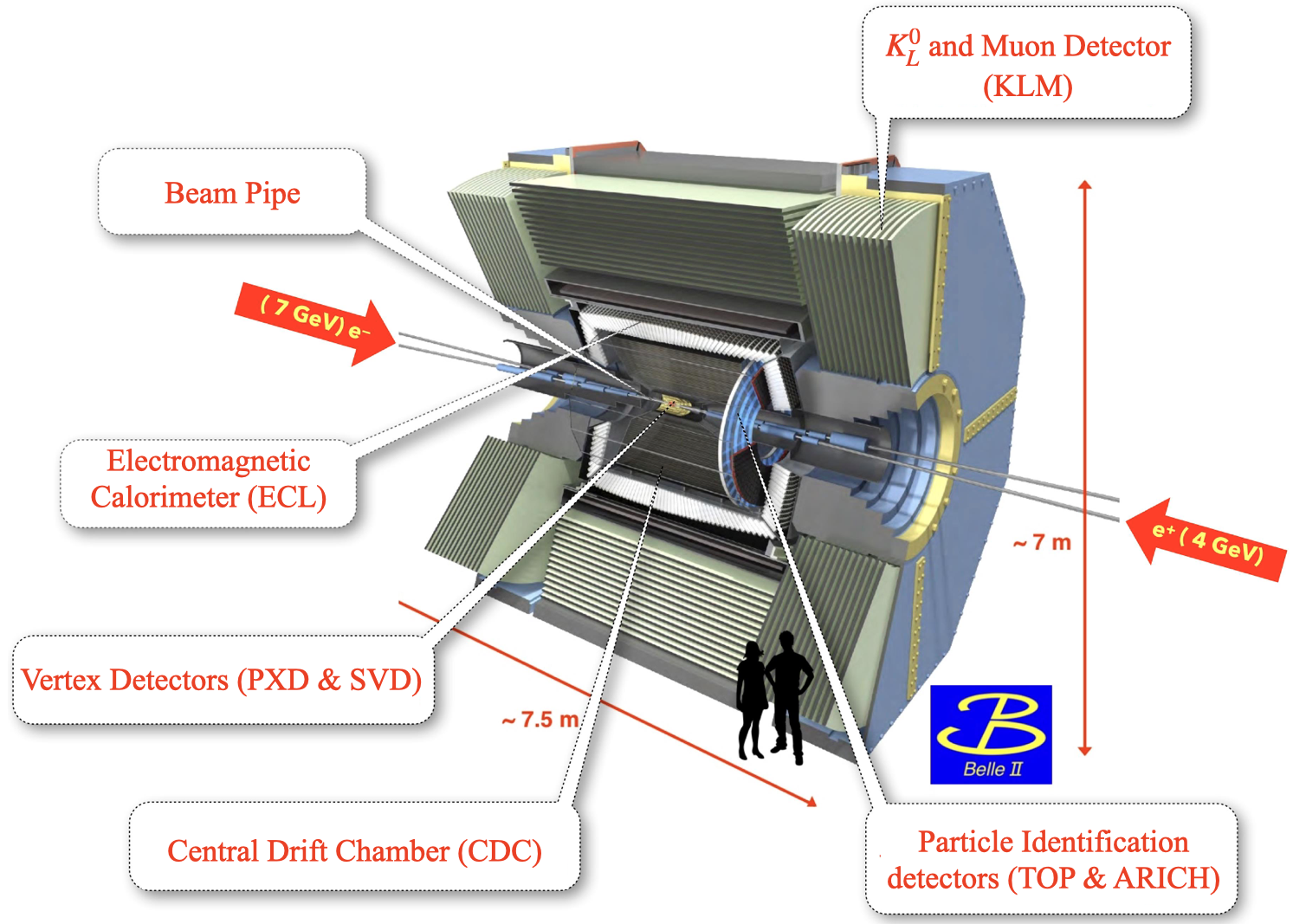}   
\caption{Schematic view of the Belle II detector~\cite{Belle-II:2010dht}. The self-annotated figure is based on overview available from SuperKEKB/Belle II public page:
\url{https://www.belle2.org/project/super_kekb_and_belle_ii} (accessed on 11 April 2022). Picture copyright and credit: KEK.}
\label{belle2det}
\end{figure}

From the IP outward, the~main components of the Belle II detector are
vertexing and tracking detectors, particle identification systems,
calorimeter and muon chambers,
as shown in Figure~\ref{belle2det}~\cite{Belle-II:2010dht}.
The tracking detectors consist of an inner Silicon PiXel Detector (PXD),
a Silicon Vertex Detector (SVD) and a Central Drift Chamber (CDC).
Two dedicated particle identification systems are
the Time-Of-Propagation (TOP) detector in the barrel region and
the Aerogel Ring-Imaging CHerenkov detector (ARICH) in the forward endcap.
These are surrounded by an Electromagnetic CaLorimeter (ECL) and
a superconducting solenoid providing a homogeneous magnetic field of 1.5 T.
A $K_L^0$ and Muon detector (KLM) is the largest
and outermost part of the Belle II~detector.

Some upgrades of the Belle II detector~\cite{Belle-II:2010dht,Belle-II:2018jsg} over the Belle detector~\cite{Belle:2000cnh} are:
\begin{itemize}
\item Vertexing:\\
In Belle, the~beam pipe was at 15~mm~\cite{Abe:2004ma},
the innermost layer of
a 4-layer silicon vertex detector~\cite{Kibayashi:2006vj}
was at 20~mm
and the outermost layer of the vertex detector was at a radius of 88~mm.
In Belle II, the~beam pipe is at 10~mm,
the inner two layers of the PXD{{, consisting of silicon pixels,}} are closer to the IP at 14~mm and 22~mm,
respectively, and~the outermost layer of the four layers of the SVD{{, consisting of silicon strips,}}
goes to a larger radius of 140~mm.
{{The PXD is}} based
on the Depleted Field Effect Transistor (DEPFET) technology,
which allows for thin sensors with 50~$\upmu$m thickness.
The readout of the new silicon strip detector  is based on the APV25 chip,
which has a much shorter shaping time  to accommodate for higher background rates
in Belle II than the VAITA chip-based readout used at Belle.
As a result of these upgrades, considerably better performance
is expected in Belle II than Belle.
For example, the~vertex resolution at Belle II is improved by
the excellent spatial resolution of the two innermost pixel detector layers.
\item Tracking:\\
The large volume CDC at Belle II,
with 56 layers organized in 9 super-layers,
has smaller drift cells than in Belle.
CDC starts just outside the expanded silicon strip detector,
and extends to a larger radius of 1130~mm in Belle II
as compared to 880~mm in Belle.
The measured spatial resolution of the CDC is about 100~$\upmu$m,
while the relative precision of the $dE/dx$  measurement
for particles  with  an  incident  angle  of  $90^\circ$ is around $12\%$.
The angular resolution achieved between tracks is $\sim$4.5~mrad.
The efficiency to reconstruct $K_S^0 \to \pi^-\pi^+$ decays in Belle II
is also improved because the silicon strip detector occupies a larger volume.
\item Particle Identification:\\
Belle II has two completely new, more compact particle identification devices
of the Cherenkov imaging type:
TOP in the barrel and ARICH in the endcap regions.
Both detectors are equipped with very fast read-out electronics,
leading to very good kaon versus pion separation in the kinematic limits
of the experiment.
The two Cherenkov detectors are designed to differentiate between
$K$ and $\pi$ particles over the entire momentum range,
and also differentiates among $\pi$, $\mu$, and~$e$ below 1~GeV/c.
\item Calorimetry:\\
The ECL is made of CsI(Tl) scintillation crystals of size 6 cm $\times$ 6 cm each
with high light output, a~short radiation length, and~good mechanical properties,
covering the range of 12$^\circ$ $<\theta<$ 155$^\circ$ in the polar angle,
e.g.,~90\% of solid angle coverage in the center-of-mass system.
The ECL is divided into two parts: the barrel and the endcap.
While the barrel part consists of 6624 crystals,
the endcap part consists of 2112 crystals.
The new electronics of the ECL are of the wave-form-sampling type,
which has particular relevance in missing-energy studies
by reducing the noise due to pile up considerably.
The ECL is able to detect neutral particles in a wide energy range,
from 20~MeV up to 4~GeV,
with a high resolution of $\sigma_E/E$ $=$ 4\% at 100~MeV,
and angular resolution of 13~mrad (3~mrad) at low (high) energies.
This gives a mass resolution for reconstructing $\pi^0\to\gamma\gamma$
of about 4.5~MeV/c$^2$~\cite{BaBar:2014omp}.
\item $K_L^0$ and Muon Detection:\\
The $K_L^0$ and muon detector (KLM) at Belle was based
on glass-electrode resistive plate chambers (RPC).
Since larger backgrounds are expected in the high luminosity environment
at Belle II, the~upgraded KLM system consists of RPC
only in some parts of the barrel.
The two innermost layers in the barrel and the entire endcap section
of KLM at Belle II consist of layers of scintillator strips
with wavelength shifting fibers,
read out by silicon photomultiplier (SiPMs) as light sensors~\cite{Balagura:2005gh}.
Although the high neutron background can cause damage to the SiPMs,
the upgraded KLM has been demonstrated to operate reliably during irradiation tests by
appropriately setting the discrimination thresholds.
\end{itemize}

\subsection{Daq Upgrade of Belle~II}

The new data acquisition (DAQ) system~\cite{Yamada:2015xjy}
meets the requirements of considerably higher event rates at Belle II.
It consists of a Level One (L1)~\cite{Iwasaki:2011za} and High Level Trigger (HLT).
The L1 trigger has a latency of 5~$\upmu$s and
a maximum trigger output rate of 30~kHz,
limited by the read-in rate of the DAQ.
The HLT must suppress online event rates to 10~kHz for offline storage
using complete reconstruction with all available information from the entire detector.
To enable readout from high-speed data transmission,
a peripheral component interconnect express based readout module (PCIe40)
with high data throughput of up to 100 Gigabytes/s was adopted
for the upgrade of the Belle II DAQ system~\cite{Zhou:2020qed}.
The trigger system at Belle II achieves almost
100~$\mathrm{\%}$ trigger efficiency for $\Upsilon(4S)\rightarrow B\bar{B}$ events
and nearly high efficiency for other physics processes of interest,
e.g., $\tau$-pair~events.


\section{Search~Strategies}
\unskip

\subsection{Event~Topology}
\label{EvtTopo}

B-Factories typically operate at center-of-mass energies
around the $\Upsilon(4S)$ resonance, e.g.,~10.58~GeV.
Tau-pair production via $e^-e^+$ annihilation in this energy regime
leads to cleanly separated event topology
associated with the decay of each $\tau$ lepton,
and are well simulated by state-of-the-art event generators:
{\texttt{KK2F}}~\cite{Jadach:1999vf,Ward:2002qq,Arbuzov:2020coe},
{\texttt{Tauola}}~\cite{Jadach:1993hs,Chrzaszcz:2016fte} and
{\texttt{Photos}}~\mbox{\cite{Barberio:1993qi,Davidson:2010ew}}.
Searches for LFV in $\tau$ decays in B factories exploit these event characteristics,
assuming that only one of the two $\tau$'s produced in the $e^-e^+\to\tau^-\tau^+$ process
could have decayed in this rare mode, and~the other $\tau$ decays
via the allowed SM processes. By~dividing the event into a pair of hemispheres
perpendicular to the thrust axis~\cite{Brandt:1964sa,Farhi:1977sg}
in the center-of-mass frame, $\tau$ decay products can thus be identified
as coming from the signal-side and the tag-side,
corresponding to decay via LFV and the SM decays of the $\tau$ lepton, respectively.

\subsection{Signal Characteristics}

The characteristic feature of $\tau$ decays via LFV is that the final state
does not contain $\nu_\tau$.
Thus, there is no missing momentum associated with the signal-side,
and the kinematics of the signal $\tau$ lepton can be completely
reconstructed from measurements of the final state particles.
Simulation studies for more than a hundred possible decays via LFV
that can be searched with such signal characteristics
for each sign of the $\tau$ lepton, are possible
with recent updates of the {\texttt{Tauola}}
event generator~\cite{Chrzaszcz:2016fte,Antropov:2019ald,Banerjee:2021rtn},
which have been seamlessly integrated into the software of the Belle II~experiment.

A very interesting feature of $\tau$-pair production in $e^-e^+$ annihilation
is that the energy of each $\tau$ lepton is known to be exactly
half of the center-of-mass (CM) energy of the collision,
except for corrections due to initial and final state radiations.
Therefore, the~uncertainty of the energy of the $\tau$ lepton
is independent of the performance of the detector,
and is known from the beam energy spread of SuperKEKB
to be approximately 5~MeV~\cite{Belle-II:2018jsg}.

As a first example of the signal mode, let us consider $\taulg$ decays,
which are predicted with rates just lower than the current experimental bounds
in the widest variety of NP models
and are hence regarded as “golden modes” in searches for LFV.
The total energy in the CM frame of the $\tau$ decay products
in the signal-side is $E_{\ell\gamma}^{\mathrm{CM}} = \sqrt{s}/2$,
and the invariant mass of the $\ell\gamma$ pair can be calculated as
$m_{\ell\gamma} = 2 \sqrt{p_\ell E_\gamma} \sin(\frac{\theta}{2})$,
where $p_\ell$ is the magnitude of the three-momentum of the lepton,
$E_\gamma$ is the energy of the photon,
and $\theta$ is the opening angle between them.
The invariant mass is ideal as a discriminating variable,
because its resolution is given by~\cite{Fabre:1992vpa}:\vspace{-6pt}
$$\frac{\Delta m_{\ell\gamma}}{m_{\ell\gamma}}
= \frac{1}{2} \sqrt{  \Big(\frac{\Delta p_\ell}{p_\ell}\Big)^2
+  \Big(\frac{\Delta E_\gamma}{E_\gamma}\Big)^2
+  \Big(\frac{\Delta \theta}{\tan \frac{\theta}{2}}\Big)^2
}
$$
which simultaneously combines all the available experimental precision
on the measured energy/momentum from the calorimeter/tracking systems
with the measured uncertainties on the position measurements of the
observable final state decays products.
The resolution of this kinematic variable is further improved
by considering the beam-energy-constrained mass,
$M_{\mathrm{bc}}$,  given as:\vspace{-6pt}
$$
M_{\mathrm{bc}}= \sqrt{(E_{\mathrm{beam}}^{\mathrm{CM}})^{2} - |\vec{p}_{\ell\gamma}^{\mathrm{~CM}}|^{2}},
$$

\noindent where $E_{\mathrm{beam}}^{\mathrm{CM}}=\sqrt{s}/2$
and $\vec{p}_{\ell\gamma}^{\mathrm{~CM}}$ is the sum of the lepton and photon momenta
in the CM frame, because~the resolution of $E_{\mathrm{beam}}^{\mathrm{CM}}$
comes from the accelerator instead of the~detector.

\begin{figure}[H]
\includegraphics*[width=.3\textwidth,height=0.222\textheight]{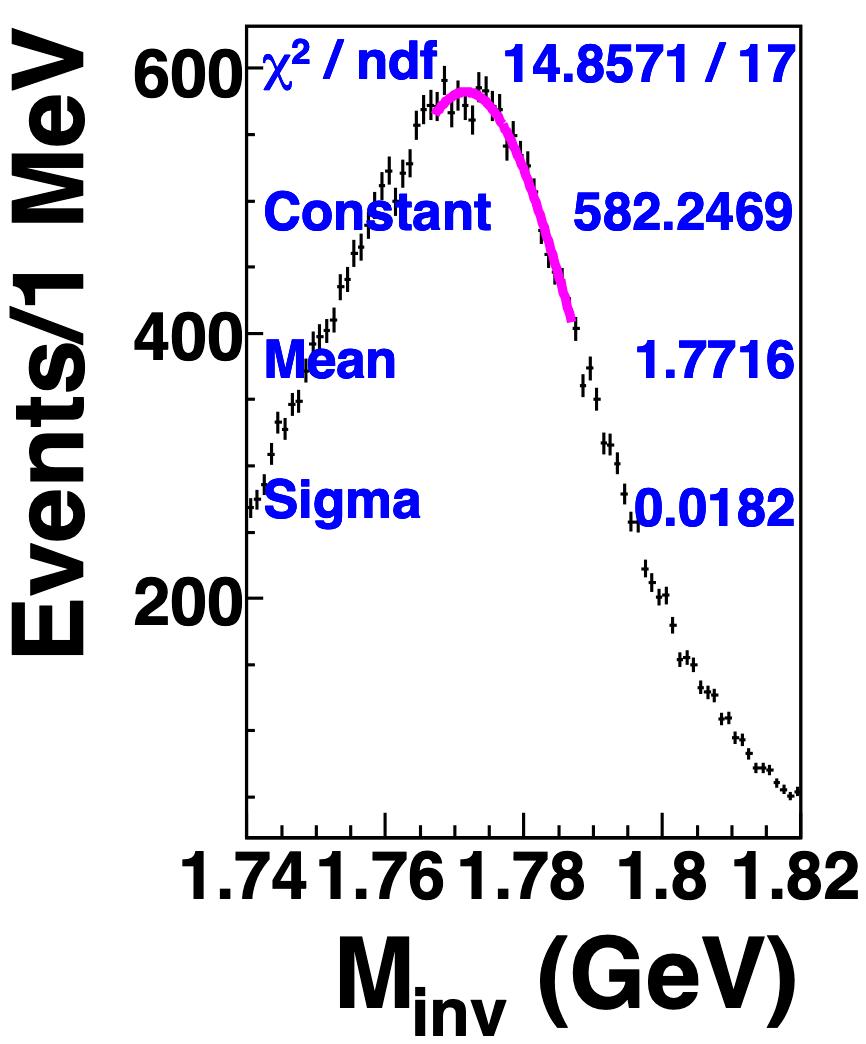}  
\includegraphics*[width=.3\textwidth,height=0.222\textheight]{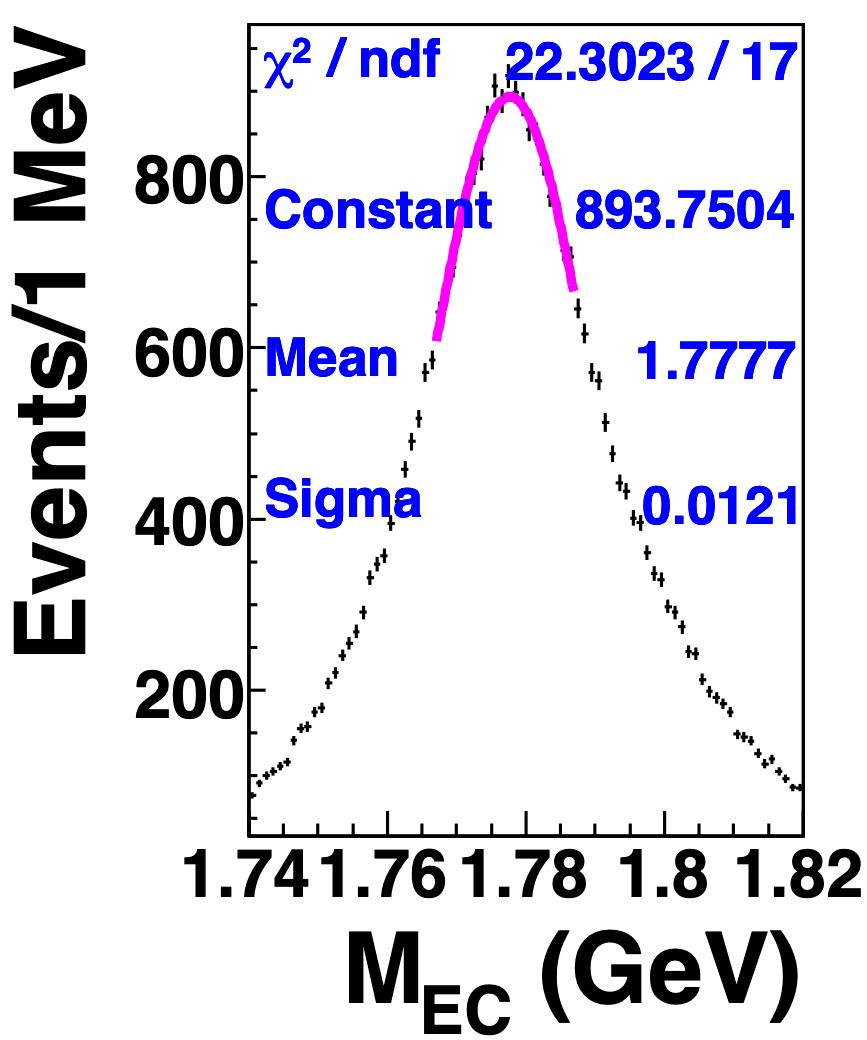}   
\includegraphics*[width=.3\textwidth,height=0.222\textheight]{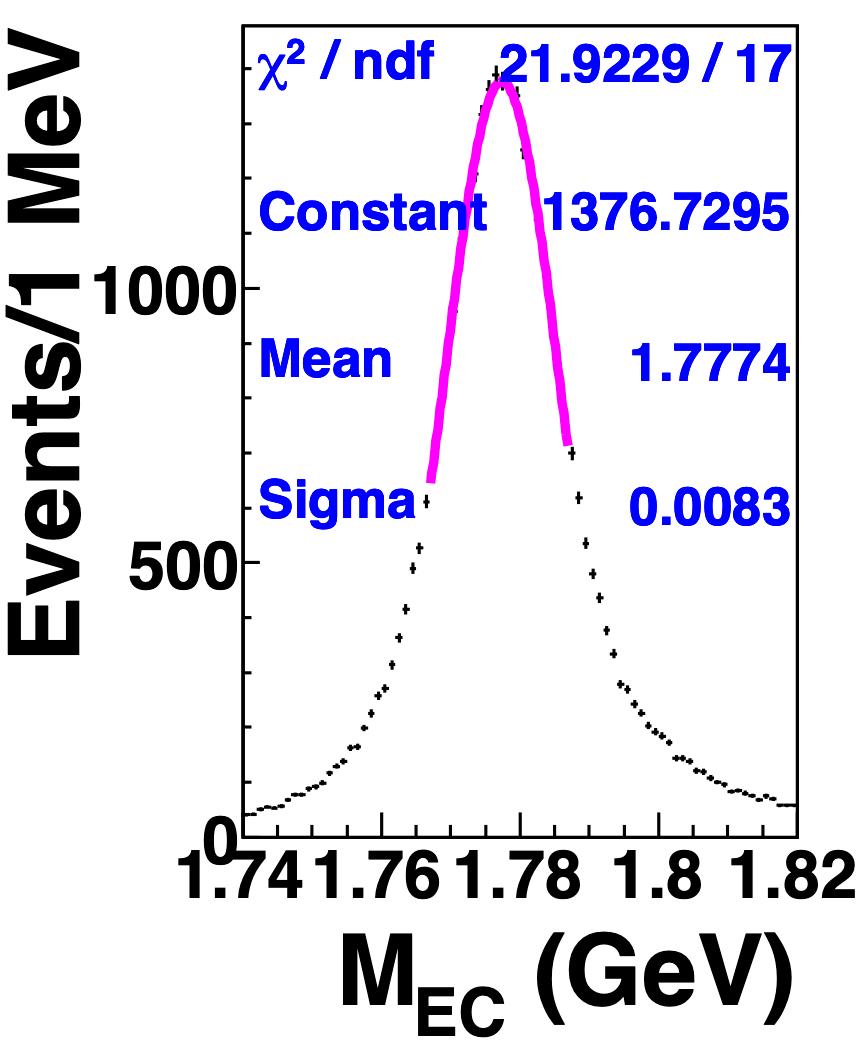}          
\vspace{-6pt}
\caption{Invariant mass (\textbf{left}), $\mec$ without vertex constraint (\textbf{middle})
and $\mec$ with vertex constraint (\textbf{right})
obtained from simulated samples during analysis strategy development studies
for $\taumg$ search at the \babar\ experiment~\cite{BaBar:2009hkt}.}
\label{fig:minv-mec-mg}
\end{figure}

The beam-energy-constrained $\tau$ mass,
labeled somewhat differently as $\mec$ for the \babar\ search~\cite{BaBar:2009hkt},
is typically obtained from a kinematic fit that constrains the CM energy of the $\tau$
to be $\sqrt{s}/2$. Its resolution was further improved in the \babar\ search
by assigning the origin of the photon candidate to the point of the closest approach
of the signal lepton track to the $e^-e^+$ collision axis.
Figure~\ref{fig:minv-mec-mg} shows a comparison study
using a simulated sample of signal $\taumg$ events,
where the resolution of the invariant mass is crudely estimated
to be 18.2~MeV/c$^{2}$ from a single Gaussian fit to the peak of the distribution,
while that of $\mec$ is 12.1~MeV/c$^{2}$,
which improves further to 8.3~MeV/c$^{2}$ after the vertex~constraint.

\begin{figure}[H]
\includegraphics*[width=.49\textwidth]{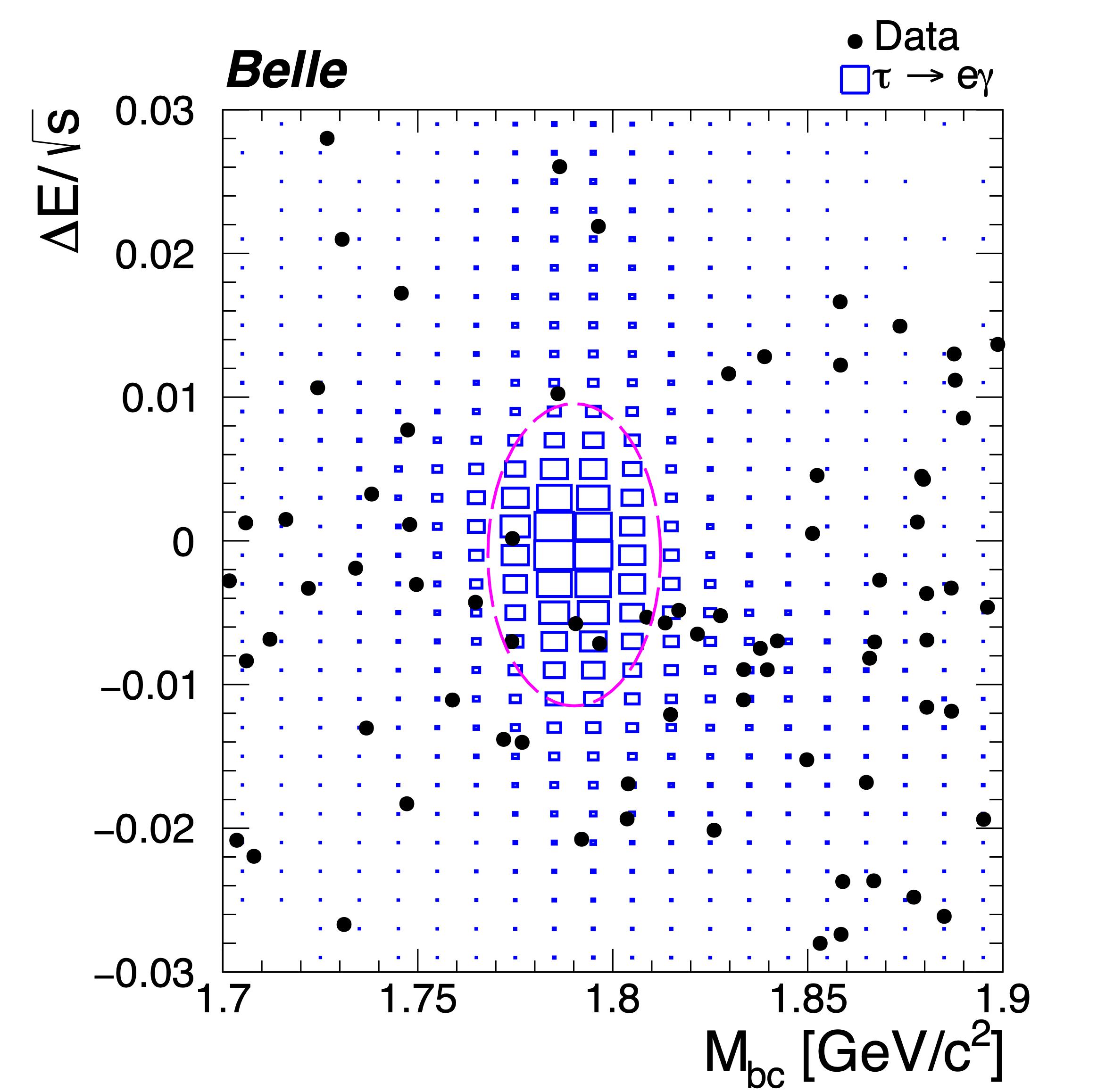}
\includegraphics*[width=.49\textwidth]{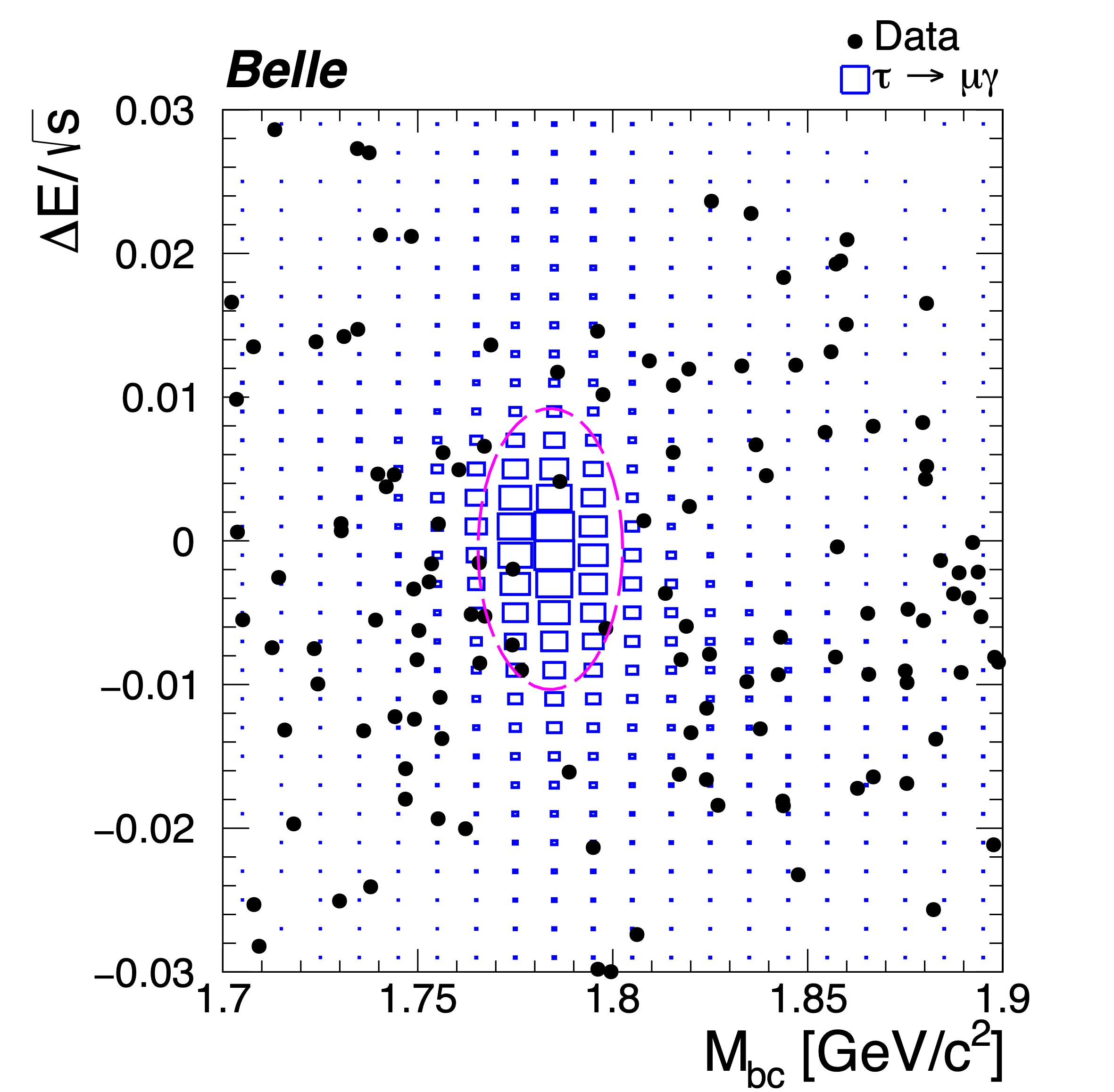}
\caption{Two-dimensional distributions of $\Delta E/\sqrt{s}$ versus $M_\mathrm{bc}$
for Belle searches~\cite{Belle:2021ysv} 
for $\taueg$ (\textbf{left}) and $\taumg$ (\textbf{right}) decays.
Black points are data, blue squares are $\taulg$ signal MC events,
and magenta ellipses show the signal region~($\pm 2\sigma$ region).
This figure has been reprinted with permission from Ref.~\cite{Belle:2021ysv}. 
}
\label{fig:belle_taulg}
\end{figure}

\begin{figure}[H]
\includegraphics*[width=.49\textwidth]{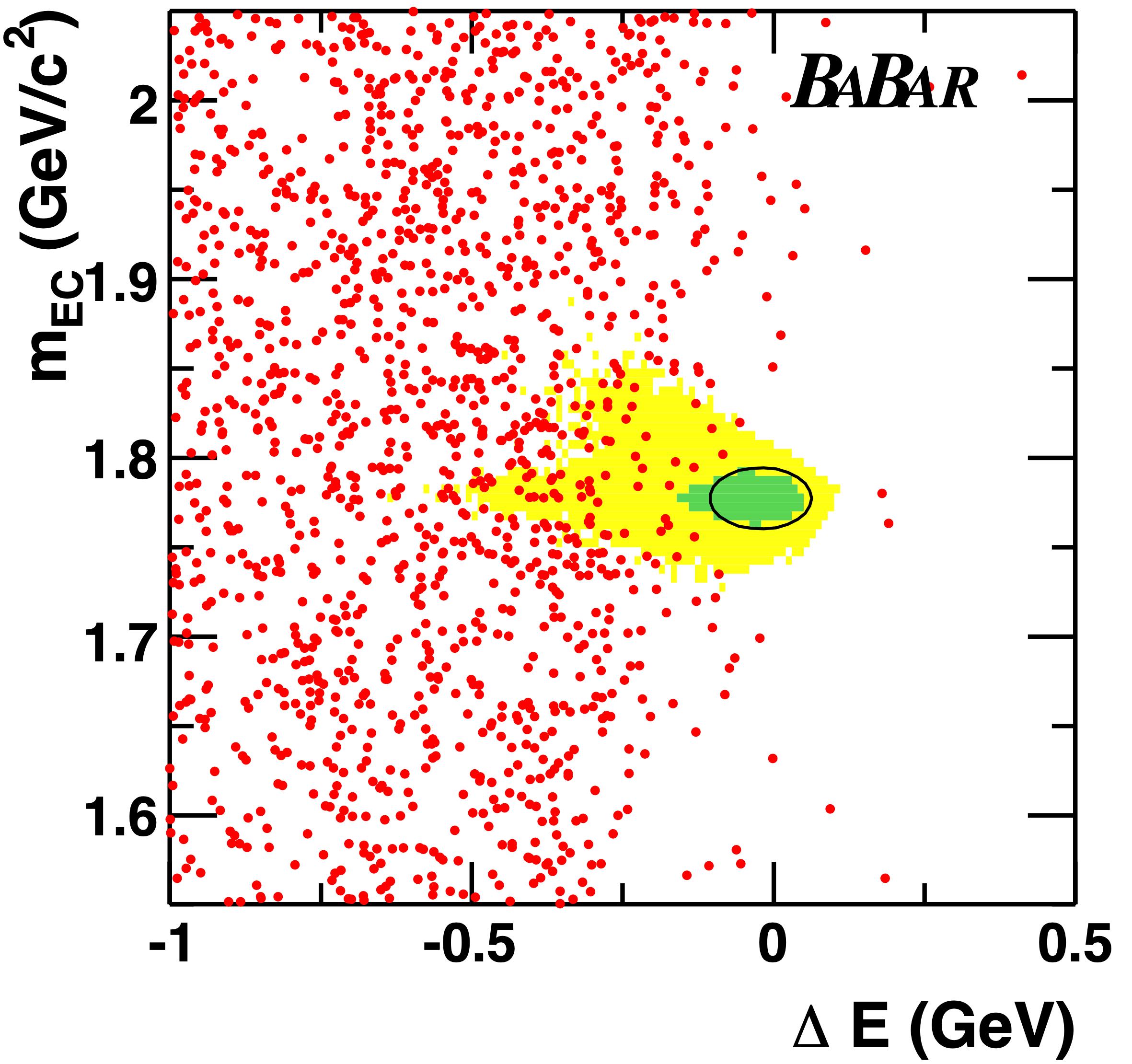}
\includegraphics*[width=.49\textwidth]{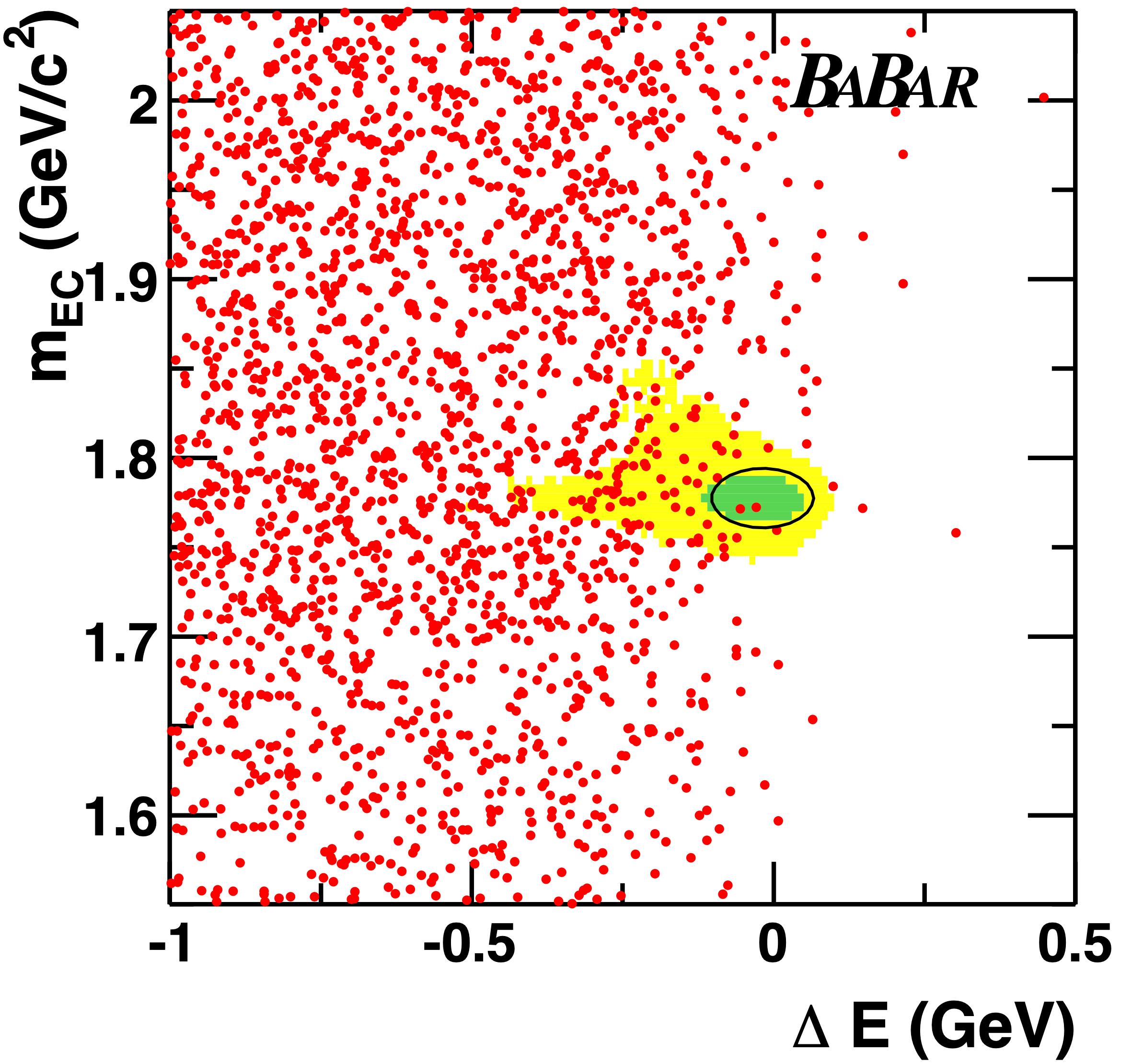}
\caption{
Two-dimensional distributions in the $\mec$ vs. $\DeltaE$ plane for the \babar\
searches~\cite{BaBar:2009hkt} 
for $\taueg$ (\textbf{left}) and $\taumg$ (\textbf{right}) decays.
Data are shown as dots and contours containing 90\% (50\%) of signal MC events
are shown as light-yellow (dark-green) shaded regions,
along with the  $2\sigma$ contours shown as black ellipses.  
This figure has been reprinted with permission from Ref.~\cite{BaBar:2009hkt}.
}
\label{fig:babar_taulg}
\end{figure}

The most distinguishing feature of signal events is obtained
by considering the characteristic mass of the decay products of the LFV $\tau$ decays
along with the normalized difference in their energy
from half the center-of-mass energy
in $e^-e^+$ annihilation,
so that the search can be uniformly performed
at energies other than the $\Upsilon(4S)$ peak,
to take advantage of the larger luminosity including all the recorded data:
$$\Delta E/\sqrt{s} = ( E_{\ell\gamma}^{\mathrm{CM}} - \sqrt{s}/2 )/\sqrt{s}.$$
The signal events are clustered around
$M_{\mathrm{bc}} \sim m_{\tau}$ and $\Delta E/\sqrt{s} \sim 0$
in the two-dimensional plots of $\Delta E$ vs. $M_{\mathrm{bc}}$,
as shown in Figures~\ref{fig:belle_taulg} and~\ref{fig:babar_taulg}
for $\taueg$ (left) and $\taumg$ (right) searches
at Belle~\cite{Belle:2021ysv} and \babar~\cite{BaBar:2009hkt} experiments, respectively,
where the variable $m_{EC}$ refers to the beam-energy-constrained mass in the latter,
as mentioned~earlier.

\unskip

All analyses are developed in a blind manner,
e.g., optimizing the event selection before
looking at the data events inside the signal region,
to avoid experimental bias in the search for LFV $\tau$ decays.
The search sensitivity can be optimized to give the smallest expected upper limit
in the background-only hypothesis inside a $2\sigma$ ellipse,
for example, amongst other possible choices.
A typical $2\sigma$ signal region is defined as the following elliptical regions:
\begin{eqnarray*} \label{eq:sr}
\frac{(M_{\mathrm{bc}} - \mu_{M_{\mathrm{bc}}})^{2}}{(2\sigma_{M_{\mathrm{bc}}})^{2}}
&+&
\frac{(\Delta E/\sqrt{s} - \mu_{\Delta E /\sqrt{s}})^{2}}{(2\sigma_{\Delta E/\sqrt{s}})^{2}} < 1.0, \\ \nonumber
\\ \nonumber
\sigma_{M_{\mathrm{bc}}} &=& 0.5(\sigma_{M_{\mathrm{bc}}}^{\mathrm{high}} + \sigma_{M_{\mathrm{bc}}}^{\mathrm{low}}), \\ \nonumber
\\ \nonumber
\sigma_{\Delta E/\sqrt{s}} &=& 0.5(\sigma_{\Delta E/\sqrt{s}}^{\mathrm{high}} + \sigma_{\Delta E/\sqrt{s}}^{\mathrm{low}}). \nonumber
\end{eqnarray*}
Here, $\sigma^{\mathrm{high/low}}_{M_{\mathrm{bc}}}$ and $\sigma^{\mathrm{high/low}}_{\Delta E/\sqrt{s}}$
are the widths on the higher$/$lower side of the peak obtained by fitting the signal distribution to an asymmetric Gaussian function~\cite{Belle:2007qih}.

\textls[-25]{For the Belle search~\cite{Belle:2021ysv}, the~resolutions are
$\sigma_{M_{\mathrm{bc}}}^{\mathrm{high/low}} = 11.55\pm0.27/10.59\pm0.19$ MeV/c$^{2}$} and
$\sigma_{\Delta E/ \sqrt{s}}^{\mathrm{high/low}} = (6.1\pm0.7)/(4.4\pm0.3)\times10^{-3}$ for $\taueg$ events;
and $\sigma_{M_{\mathrm{bc}}}^{\mathrm{high/low}} = 11.08\pm0.08/7.46\pm0.23$ MeV/c$^{2}$ and
$\sigma_{\Delta E/ \sqrt{s}}^{\mathrm{high/low}} = (5.6\pm0.4)/(4.2\pm0.2)\times10^{-3}$ for $\taumg$ events.
The mean values are
$\langle {M_{\mathrm{bc}}} \rangle = 1.79$ MeV/c$^{2}$
and $\langle {\Delta E/\sqrt{s}} \rangle =  -1.0\times10^{-3}$ for $\taueg$ events,
and
$\langle {M_{\mathrm{bc}}} \rangle = 1.78$ MeV/c$^{2}$ and $\langle {\Delta E/\sqrt{s}} \rangle  = -0.6\times10^{-3}$ for $\taumg$ events.
For the \babar\ search~\cite{BaBar:2009hkt},
the resolutions for $\mec$ and $\DeltaE$ are 8.6 MeV/c$^{2}$ and \mbox{42.1 \mev} for $\taueg$,
and 8.3 MeV/c$^{2}$ and 42.2 \mev for $\taumg$, respectively,
centered on 1777.3 MeV/c$^{2}$ and $-$21.4 \mev for $\taueg$,
and 1777.4 MeV/c$^{2}$ and $-$18.3 \mev for $\taumg$.
Shifts from zero for $\langle \DeltaE \rangle$
are mostly due to initial and final state~radiations.

The mass and energy kinematic variables typically have a small correlation
arising from initial and final state radiation,
as well as energy/momentum scale calibration effects.
For the \babar\ search~\cite{BaBar:2009hkt}, the~correlation was estimated to be
$-$8.5\% and $-$8.4\% for the $\taueg$ and $\taumg$ decays, respectively,
around the core region.
Without the beam-energy constraint, the~correlation between the invariant mass
and energy variables are typically much higher.

LFV process in $\tau$ decays containing a resonance in the final state
are identified by the presence of a peak in the invariant mass of the daughter particles
in the simulation of the signal process.
For example, distributions of invariant mass of the
$\pi^+\pi^-$, $K^+K^-$, $\pi^+\pi^-\pi^0$, $K^+\pi^-$, $\pi^+K^-$ and $\pi^+\pi^-$ systems
in the signal-side are studied {{and confirmed to contain the respective resonances}} in searches for
$\tau^- \to \mu^- \rho^0$,
$\tau^- \to \mu^- \phi$,
$\tau^- \to \mu^- \omega$,
$\tau^- \to \mu^- K^{\ast 0}$,
$\tau^- \to \mu^- \bar{K}^{\ast0}$ and
$\tau^- \to \mu^- f_0(980)$ decays, respectively,
performed by the Belle experiment~\cite{Belle:2011ogy, Belle:2008pdf}.
The selected mass regions ensure that the signal is unambiguously selected in the corresponding~searches.

\subsection{Background Suppression}

Background events containing leptons from decays of heavy quarks
are easily suppressed by appropriate cuts on Fox-Wolfram moments~\cite{Fox:1978vu},
and on the invariant mass of all decay products on the tag-side.
The characteristic difference between $\tau$-pairs events with LFV decays
and backgrounds consisting of generic $\tau$-pair, di-lepton,
two-photon production and $q\bar{q}$ processes
{{(where $q = u,~ d \mathrm{~or~} s$),}}
in the number of neutrinos in the signal-side and tag-side,
as defined by the event topology in Section~\ref{EvtTopo},
are shown in Table~\ref{NumNeutrino}.

\begin{table}[H]
\setlength{\tabcolsep}{6.2mm}

\caption{Number of neutrinos in the event for signal and background processes.}
\label{NumNeutrino}
\begin{tabular} {cccc}
\toprule
\textbf{ \# }\textbf{of }\boldmath{$\nu$}\textbf{'s} & \textbf{LFV Decays }& \textbf{Generic} \boldmath{$\tau$}\textbf{-Pair} & \textbf{Other Backgrounds} \\ \midrule
Signal-side   & 0 & 1{--}2 & 0\\
Tag-side      & 1{--}2 & 1{--}2 & 0\\\bottomrule
\end{tabular}
\end{table}

Since decay products of the $\tau$ decay via LFV in the signal-side
do not contain any neutrino, the~direction of the $\tau$ lepton
in the tag-side can be precisely obtained in the center-of-mass frame
by reversing the total momentum of the signal-side.
This allows for good kinematic reconstruction of the missing mass in the tag-side,
assuming that in the CM frame, the~tag-side $\tau$ momentum is opposite that of
the signal-side $\tau$ momentum and that its energy is constrained to be
half the center-of-mass energy.
Thus, {{selection of events with small values of}} the square of the missing mass ($m_\nu^2$) in the tag-side
play an important role in the suppression of the background events~\cite{Belle:2021ysv}.

Additional selection criteria are also used to suppress
the backgrounds in the different LFV decay modes,
which are mostly accidental in nature, except~in $\taueg$ and $\taumg$ searches.
The dominant background in the searches arise from $\tau^{+}\tau^{-}$ events decaying
via the $\tau^{\pm}\rightarrow e^{\pm}\nu_{e}\nu_{\tau}$
~($\tau^{\pm}\rightarrow\mu^{\pm}\nu_{\mu}\nu_{\tau}$)
channel with a photon coming from initial-state radiation or beam background.
The $e^{+}e^{-}\gamma$ and $\mu^{+}\mu^{-}\gamma$ events are subdominant,
and are estimated to contribute to  <5\% of the total backgrounds
in the Belle search~\cite{Belle:2021ysv}.
Contributions from other sources of backgrounds,
such as two-photon and $q\bar{q}$ processes,
are estimated to be quite small in the signal~region.

\begin{figure}[H]
\includegraphics*[width=\textwidth,height=.2\textheight]{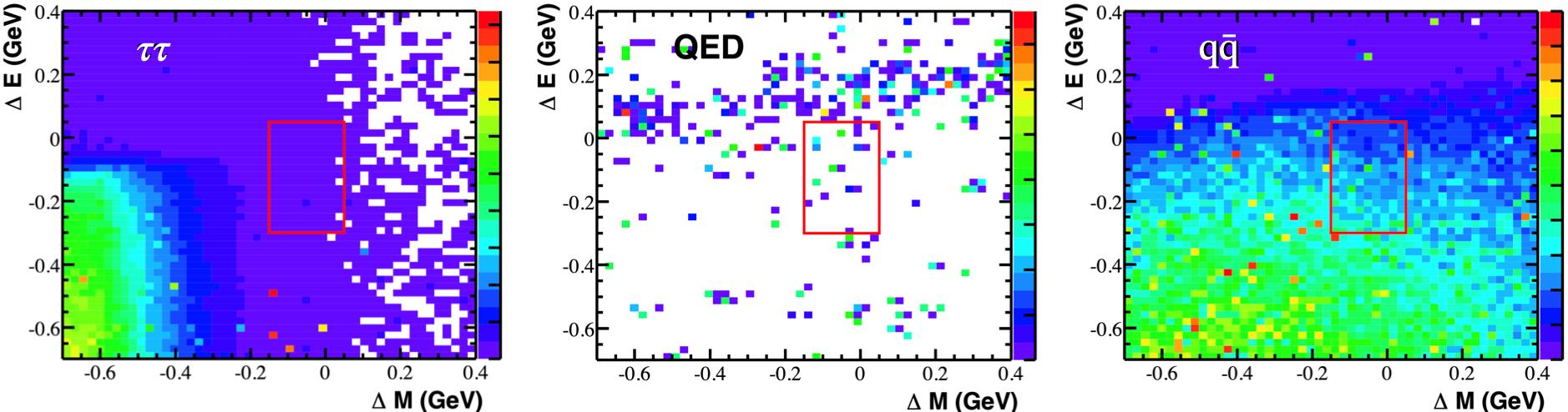}
\caption{Two-dimensional distributions for $\tau\tau$, QED, and~$q\bar{q}$ background processes in the $(\Delta M, \DeltaE)$ plane
obtained from simulated samples during analysis strategy development studies
for $\taummm$ search at the \babar\ experiment~\cite{BaBar:2010axs}.
Rectangular box-shaped regions mostly populated by the signal processes are shown in red.}
\label{fig:taulll_bkg_dmde}
\end{figure}

Furthermore, each component of the background processes has distinctive features
as visible in their respective two-dimensional distributions
in the $(\Delta M, \DeltaE)$ plane,
where $\Delta M$ denotes the difference between the characteristic mass
of the system of $\tau$-daughters and the well-known mass of the $\tau$-lepton
= $(1776.86 \pm 0.12)\mev$~\cite{PDG2020},
and $\DeltaE$, as defined above. The~shapes of the leading backgrounds
in search of $\taummm$ decays
as performed at the \babar\ experiment~\cite{BaBar:2010axs}
are shown in Figure~\ref{fig:taulll_bkg_dmde},
where the red box indicates the rectangular boundaries of a generic region
mostly populated by the signal processes. The~SM $\tau\tau$ background events
are generally restricted to {{small}} negative values of both $\Delta M$ and $\DeltaE$ variables,
because the {{reconstruction of signal}} event topology does not account for the {{neutrinos present in SM $\tau$ decays}}.
QED background events are mostly dominated by di-lepton production
as the main underlying hard process and typically lie within a narrow horizontal band
across the $\Delta M$ variable centered around slightly positive values of $\DeltaE$,
due to the presence of a pair of extra charged particles in such events.
The QCD background events from various $q\bar{q}$ processes
tend to populate the plane uniformly
across the $\Delta M$ variable and drop towards large values of the $\DeltaE$ variable.
The expected background rates inside the signal region can be obtained
by fitting the observed data in the $(\Delta M, \DeltaE)$ plane
to a sum of probability density functions. Such data-driven estimates,
based on the shapes predicted by respective simulation samples
and validated by data-driven control regions, scale well with larger data statistics.
Thus, the~background uncertainties can be controlled in a statistical manner,
which is very useful in rare searches with high luminosity data~sets.

\subsection{Upper Limit Estimation}

No excess of events has ever been observed in searches for LFV in $\tau$ decays.
The upper limit at 90\% confidence level (CL) on signal branching fraction ($\BR^{90}_{UL}$) is calculated as:
$$\BR^{90}_{UL} = \frac{S^{90}_{UL}}{2N_{\tau\tau} \eff},$$
where
$N_{\tau\tau}$ is the number of $\tau$-pairs produced,
$\eff$ is the reconstruction efficiency of the signal decay mode,
and $S^{90}_{UL}$ is the 90\% CL upper limit on number of signal events.
The factor of two enters into the denominator because either one of the two $\tau$ leptons produced in the event
can decay into the rare signal channel coming from~LFV.

\subsubsection{Number of $ \tau$-Pairs~Produced}

$N_{\tau\tau}$ is obtained by summing over
the product of luminosity $({\cal{L}})$ and the $\tau$-pairs production cross-section $(\sigma_{\tau\tau})$
at each of the center-of-mass energies ($\sqrt{s}$) where the search is conducted.
The cross-section of $\tau$-pair production in $e^-e^+$ annihilation follows a characteristic ${1/s}$ dependence,
but receives additional contribution from decays of the $\Upsilon(nS)$ resonances for $n = 1, ~2 {~\rm{and}~} 3$,
according to the known branching fractions~\cite{PDG2020}:
${\cal{B}}(\Upsilon (1S) \to \tau^-\tau^+) = (2.6 \pm 0.1) \%$,
${\cal{B}}(\Upsilon ({{2}}S) \to \tau^-\tau^+) = (2.0 \pm 0.2) \%$ and
${\cal{B}}(\Upsilon ({{3}}S) \to \tau^-\tau^+) = (2.3 \pm 0.3) \%$.
Beyond the open-beauty threshold,
contributions from resonances with $n\geq 4$ are~negligible.

The Belle experiment collected data with center-of-mass energies around the peak of
$\Upsilon (nS)$ resonances corresponding to luminosities of $5.7\invfb$, $24.9\invfb$ and $2.9\invfb$ at $n = 1, ~2 {~\rm{and}~} 3$, respectively,
while the \babar\ experiment collected luminosities of $13.6\invfb$ and $28.0\invfb$
at center-of-mass energies corresponding to the $\Upsilon (nS)$ peak
at $n = 2{~\rm{and}~} 3$, respectively,
as reported in Table~3.2.1 in the reference~\cite{BaBar:2014omp}.
The statistical errors on these measured luminosities are much smaller than
the systematic errors, which are estimated to be 1.4\% at the Belle experiment,
and 0.7\% (0.6\%) at the \babar\ experiment for $n = 2{~\rm{and}~} 3$, respectively~\cite{BaBar:2014omp}.

The numbers of $\Upsilon (nS)$ produced in the Belle experiment are
$(102 \pm 2) \times 10^6$,
\mbox{$(158 \pm 4)$} $ \times 10^6$ and
$(11 \pm 0.3) \times 10^6$
for $n = 1, ~2 {~\rm{and}~} 3$, respectively,
while in the \babar\ experiment, the numbers are
$(98.3 \pm 0.9) \times 10^6$ and
$(121.3 \pm 1.2) \times 10^6$
for $n = 2{~\rm{and}~} 3$, respectively,
as obtained from Table~3.2.2 in the reference~\cite{BaBar:2014omp}.
By dividing the numbers of $\Upsilon (nS)$ resonances produced
with its corresponding luminosity,
the resonant production cross-sections are estimated
for $\Upsilon(1S)$ to be $(17.89 \pm 0.43)~nb$ in the Belle experiment,
for $\Upsilon(2S)$ to be $(6.35 \pm 0.18)~nb$ and $(7.23 \pm 0.08)~nb$
in the Belle and \babar\ experiments, and~ for $\Upsilon(3S)$ to be $(3.79 \pm 0.12)~nb$ and $(4.33 \pm 0.05)~nb$
in the Belle and \babar\ experiments, respectively.
For the purposes of averaging the measured values, the~observed differences at $n = 2{~\rm{and}~} 3$ resonances are accounted for
by calculating the PDG-style scale factors~\cite{PDG2020} $S=\sqrt{\chi^2/(N-1)}$ equal to 4.38 and 4.24, respectively,
with $N=2$. Thus, the~average values are estimated to be
$(7.08 \pm 0.33)~nb$ for $\Upsilon(2S)$ and $(4.24 \pm 0.20)~nb$ for $\Upsilon(3S)$.

The total $\sigma_{\tau\tau}$ is obtained by adding the contributions
to $\tau$-pairs production from the continuum~\cite{Banerjee:2007is}
and the contributions from the decays of the $\Upsilon(nS)$ resonances
for $n = 1, ~2 {~\rm{and}~} 3$.
The estimated values of total $\sigma_{\tau\tau}$
at the peak of $\Upsilon(nS)$ resonances
and at $60\mev$ below the corresponding resonances (labelled with a “—off”)
are listed in Table~\ref{tab:cross}.

\begin{table}[H]

\setlength{\tabcolsep}{9mm}

\caption{$\sigma_{\tau\tau}$ at different center-of-mass energies corresponding to data-taking at the~B-Factories.}
\label{tab:cross}
\begin{tabular} {lclc}

\noalign{\hrule height 1pt}

\boldmath{$\sqrt{s}$} & \boldmath{$\sigma_{\tau\tau}$  $(nb)$} & \boldmath{$\sqrt{s}$} & \boldmath{$\sigma_{\tau\tau}$ $(nb)$}  \\
\midrule
$\Upsilon(1S)$—off     & $1.142 \pm 0.004$ &   $\Upsilon(1S)$         & $1.593 \pm 0.021$ \\
$\Upsilon(2S)$—off     & $1.026 \pm 0.004$ &   $\Upsilon(2S)$         & $1.157 \pm 0.017$ \\
$\Upsilon(3S)$—off     & $0.966 \pm 0.003$ &   $\Upsilon(3S)$         & $1.052 \pm 0.014$ \\
$\Upsilon(4S)$—off     & $0.928 \pm 0.003$ &   $\Upsilon(4S)$         & $0.919 \pm 0.003$ \\
$\Upsilon(5S)$—off     & $0.881 \pm 0.003$ &   $\Upsilon(5S)$         & $0.873 \pm 0.003$ \\
\noalign{\hrule height 1pt}

\end{tabular}

\end{table}
\unskip

\subsubsection{Efficiency of Signal~Reconstruction}
{{\label{sec:signal_eff}}}

The signal reconstruction efficiency receives multiplicative reduction factors
corresponding to the application of trigger, acceptance, and~event topology requirements,
particle identification criteria, background suppression and
choice of the signal region in the two-dimensional plane
given by mass versus the normalized difference of energy of the $\tau$ decay products
and $\sqrt{s}/2$.
At Belle and \babar, the~signal efficiencies were estimated to lie approximately
between $2\%$ to $12\%$, depending on the different decay channels.
For example, the~overall signal efficiency estimated
for the search for $\tau^-\to e^- \eta^\prime$ reconstructed
via the $\eta^\prime\to\rho(\to\pi^-\pi^+)\gamma$ and $\eta^\prime\to\pi^-\pi^+\eta(\to\gamma\gamma))$ decay modes
is $(0.294 \times 1 \times 4.76 + 0.445 \times 0.3943 \times 4.27)\% = 2.1\%$~\cite{Belle:2007cio},
while the search for $\tau^-\to\mu^+e^-e^-$ decays have an efficiency of 11.5\%~\cite{Hayasaka:2010np}.
In the Belle II experiment, an~increase in the signal efficiency can be expected due to higher trigger efficiencies,
improvements in the vertex reconstruction, charged track and neutral meson reconstructions, and~particle identification.
Refinements in the analysis techniques will produce a more accurate understanding
of the physics backgrounds and would thus contribute to an increase in the signal detection efficiency, which directly translates into higher sensitivities
in searches for~LFV.

\subsubsection{Upper Limit on the Number of Signal~Events}

In the case of searches with very low counts,
the search becomes a single-bin counting experiment
following a Poisson probability distribution,
with the mean count given by the expected
number of background events ($b$)
and possibly some signal events ($s$).
The likelihood function $({\cal {L}})$ is thus described by:
$$ {\cal {L}}(s,b) = \frac{e^{-(s+b)}(s+b)^N}{N!},$$
where $N$ is the number of observed events.
If the experimental resolution of the discriminating variables allows multiple bins,
the difference in shapes of the discriminating variables between the signal and background distributions
can be exploited in an extended unbinned maximum likelihood fit using:
$$
{\cal {L}}(s,b)=\frac{e^{-(s+b)}} {N!}
\prod_{i=1}^{N} (s\cdot PDF^{sig}_i + b \cdot PDF^{bkg}_i) ,
$$

\noindent where $i$ indicates the $i$-th event, $PDF^{sig}$  and $PDF^{bkg}$
are the probability density functions (PDF) for signal and sum of all background process, respectively.

$S^{90}_{UL}$ is obtained by considering
$ {\cal {L}}(s) =  \int_0^\infty {\cal {L}}(s,b) db$,
and integrating the likelihood ${\cal {L}}(s)$ up to the value that
includes 90\% of the  total integral of the likelihood function,
following a flat prior Bayesian prescription~\cite{Helene:1982pb}.
Alternatively, following a Frequentist prescription~\cite{Narsky:1999kt},
a toy Monte Carlo approach is used to generate numerous samples
with sizes that follow a Poisson distribution about the mean value
being given by the number of observed events.
Each sample is then fitted to obtain the number of signal and background events using
the same extended unbinned maximum likelihood fit procedure
as that applied to the data.
$S^{90}_{UL}$ is obtained by varying the true branching fraction of the signal
such that 90\% of the samples
yield a fitted number of signal events greater than the number of signal events in the observed data sample.
In the unified approach for finding confidence levels~\cite{Feldman:1997qc},
the order of samples in the acceptance interval
for a specific value of the number of signal events
follows an ordering principle based on likelihood ratios,
where the denominator is determined by the best fit value in each~sample.

In order to have an unbiased estimate of the expected sensitivity, a~blinding procedure should be followed to predict the expected background rate
inside the signal region (SR),
which does not depend on the observed data inside a blinding region (BR),
defined as a part of a broad fit range (FR), but~hiding data events inside the SR.
For well-controlled modeling of the total background PDF,
the number of expected background events
($N^{bkg}_{SR}$) inside the SR can then be estimated directly
from the data $N^{data}_{FR-BR}$ outside the blinded region using the formula:
$$ N^{bkg}_{SR} = \frac{\int_{SR} PDF_{bkg}}
{\int_{FR-BR} PDF_{bkg}}
\times N^{data}_{FR-BR}$$
where $\int_{SR}PDF_{bkg}$ and $\int_{FR-BR}PDF_{bkg}$
are the integrals of the background probability density functions
over the signal region and the non-blinded parts of the fit~region.

\subsection{Systematic~Uncertainties}

In terms of the number of $\tau$ decays being studied,
$t = 2 N_{\tau\tau} \epsilon$,
the number of signal events are written as $(s = \mu t)$,
where $\mu$ is the branching fraction of the signal process
and the normalization factor ${t}$ includes uncertainities on
luminosity, cross-section and the signal efficiency.
The upper limit $\BR^{90}_{UL}$ including all systematic effects
using the technique of Cousins and Highland~\cite{Cousins:1991qz},
is calculated by propagating all the measured uncertainties
onto the number of signal events $(s)$ and background events $(b)$.

Implementation of systematic effects
in the POLE (POisson Limit Estimator) program~\cite{Conrad:2002kn}
is based on the following likelihood function,
which is a convolution of a Poisson distribution
with two Gaussian resolution functions corresponding
to the signal normalization factor and background,
as described by the following formula:
$$
{{\cal L}}((\mu,t,b) = {(\mu t + b)^N e^{-(\mu t + b)} \over N!}
\frac{1}{2\pi\sigma_t\sigma_b}
e^{-{1\over 2}\left({\hat{t} - t\over\sigma_t}\right)^2
-{1\over 2}\left({\hat{b} - b\over\sigma_b}\right)^2},
$$
where $\hat t$ and $\hat b$ are the average estimates corresponding
to measured uncertainties of $\sigma_t$ and $\sigma_b$, respectively.

In searches for rare processes such as LFV in $\tau$ decays,
often a very small number of events are expected in the signal region.
Sometimes the sensitivity of the search cannot easily distinguish
a very small number of signal events
from the background-only hypothesis, and~inappropriately tends
to exclude an unusually small signal value.
To overcome such difficulties, the~upper limits can be calculated
using the $CL_s$ method~\cite{Junk:1999kv, Read:2002hq},
where the $CL_s$ is defined as the ratio of confidence levels
for the signal-plus-background hypothesis
normalized by the confidence level for the background-only hypothesis.
Asymptotic calculations of the likelihood ratios used as the test statistic
in such methods allow for a computationally efficient estimate
of the $CL_s$ intervals~\cite{Cowan:2010js}.

A Neyman construction~\cite{Neyman:1937uhy} of  $CL_s$ upper limits
including systematic uncertainties is provided by the \texttt{HistFactory} implementation~\cite{Moneta:2010pm,Cranmer:2012sba},
based on the likelihood function $\mathcal{L}(\mu,\theta_j)$ defined as:
$$
\mathcal{L}(\mu, \theta_j) =
\prod_{\textrm{channel~}} \prod_{\textrm{category~}}
\left[ \prod_{i}
\text{Poisson}(N_i|\mu \cdot t_i + b_i)
\prod_{j}
\text{Lognormal}(\theta_j|0,1)
\right],
$$
where $N_i$ is the number of events observed in the $i^{th}$ bin with
signal normalization factor and background predictions given by $t_i$ and $b_i$,
respectively, of~a multicategorical search describing,
for example, different tag-side decay modes each with different sensitivity
over possibly multiple decay channels of the signal mode.
The systematic uncertainties are constrained by nuisance parameters
$\theta_j$ corresponding to various scale factors
as determined from dedicated calibration constants of efficiency measurements
and are obtained from simulation studies or analysis of control regions in the~data.

The  \texttt{HistFactory} allows for the calculation of upper limits in both
Bayesian and Frequentist interpretations~\cite{Cranmer:2021urp},
with slightly different treatments of the nuisance parameters.
While in the former interpretation, the~nuisance parameters
are eliminated by marginalizing the posterior density,
using, for~example, Markov Chain Monte Carlo integration, 
in the latter interpretation, the~nuisance parameters
are determined by profiling the likelihood function
based on auxiliary measurements, such as control regions,
side-bands, or~dedicated calibration measurements.
Some uncertainties arising from theoretical calculations
or ad hoc estimates are not statistical in nature
and thus are not associated with auxiliary measurements.
However, log-normal probability density functions of nuisance parameters
are used to constrain all the uncertainties, by~convention.

Bayesian limits can also be calculated using the Bayesian Analysis Toolkit~\cite{Caldwell:2008fw}.


\section{Current Status and Future~Prospects}

Summary of observed limits obtained by CLEO, \babar, Belle, ATLAS, CMS, and~LHCb experiments~\cite{HFLAV:2019otj}
are shown in Table~\ref{tab:TauDecays} and Figure~\ref{fig:TauLFV}, along with
projections for two illustrative scenarios of luminosity ${\cal{L}}$ = 5~\invab and 50~\invab at the Belle II experiment~\cite{Belle-II:2022cgf}.
Projections are extrapolated from expected limits obtained at the Belle experiment.
The expected limits for $\tau^- \to \ell^-\gamma$ decays are obtained from Ref.~\cite{Belle:2021ysv}.
We assume the presence of irreducible backgrounds for $\tau^- \to \ell^-\gamma$ decays,
thus approximating the sensitivity to upper bounds
as proportional to $1/\sqrt{\cal{L}}$.
{{Given the expected number of background events in each channel from the previous searches at the Belle experiment and the improvements listed in Section~\ref{sec:signal_eff},
the background expectations corresponding to the integrated luminosity at Belle II for all other modes are still of the order of unity or less.
For such accidental backgrounds, the~sensitivity for upper bounds is proportional to $1/{\cal{L}}$, as~discussed in Section~8.2.1.3 of Ref.~\cite{Raidal:2008jk}.
The projections for the corresponding upper limits at Belle II are estimated using the Feldman and Cousins approach~\cite{Feldman:1997qc}.}

\begin{table}[H] 
\caption{Current status of observed (obs) and expected (exp) upper limits (UL)~\cite{PDG2020,Belle-II:2022cgf}. 
} \label{tab:TauDecays}
\begin{adjustwidth}{-\extralength}{0cm}
\setlength{\cellWidtha}{\fulllength/7-2\tabcolsep-0in}
\setlength{\cellWidthb}{\fulllength/7-2\tabcolsep-0in}
\setlength{\cellWidthc}{\fulllength/7-2\tabcolsep-0in}
\setlength{\cellWidthd}{\fulllength/7-2\tabcolsep-0in}
\setlength{\cellWidthe}{\fulllength/7-2\tabcolsep-0in}
\setlength{\cellWidthf}{\fulllength/7-2\tabcolsep-0in}
\setlength{\cellWidthg}{\fulllength/7-2\tabcolsep-0in}
\scalebox{1}[1]{\begin{tabularx}{\fulllength}{>{\PreserveBackslash\centering}m{\cellWidtha}>{\PreserveBackslash\centering}m{\cellWidthb}>{\PreserveBackslash\centering}m{\cellWidthc}>{\PreserveBackslash\centering}m{\cellWidthd}>{\PreserveBackslash\centering}m{\cellWidthe}>{\PreserveBackslash\centering}m{\cellWidthf}>{\PreserveBackslash\centering}m{\cellWidthg}}

\noalign{\hrule height 1pt}

\multicolumn{1}{c}{\textbf{}} & \multicolumn{3}{c}{\textbf{Observed Limits}} & \multicolumn{3}{c}{\textbf{Expected Limits}} \\ \midrule
\boldmath{$\tau^-\to$}            &  \textbf{Experiment}                   & \textbf{Luminosity}       & \textbf{UL (obs)}& \textbf{Experiment} & \textbf{Luminosity}  & \textbf{UL (exp)}        \\ \midrule

$e^-\gamma$            & Belle~\cite{Belle:2021ysv}   & 988~\invfb & 5.6$~\times~10^{-8}$ & Belle II~\cite{Belle-II:2022cgf}   & 50 \invab & 9.0$~\times~10^{-9}$  \\
& \babar~\cite{BaBar:2009hkt}  & 516~\invfb & 3.3$~\times~10^{-8}$ &                            &           &                     \\
& CLEO~\cite{CLEO:1996sqd}     & 4.68~\invfb& 2.7$~\times~10^{-6}$ &                            &           &                     \\\midrule
$\mu^-\gamma$	       & Belle~\cite{Belle:2021ysv}   & 988~\invfb & 4.2$~\times~10^{-8}$ & Belle II~\cite{Belle-II:2022cgf}   & 50 \invab & 6.9$~\times~10^{-9}$  \\
& \babar~\cite{BaBar:2009hkt}  & 516~\invfb & 4.4$~\times~10^{-8}$ &                            &           &                     \\
& CLEO~\cite{CLEO:1999lvl}     & 13.8~\invfb& 1.1$~\times~10^{-6}$ &                            &           &                     \\\midrule
$e^-\pi^0$	       & Belle~\cite{Belle:2007cio}   & 401~\invfb & 8.0$~\times~10^{-8}$ & Belle II~\cite{Belle-II:2022cgf}   & 50~\invab & 7.3$~\times~10^{-10}$ \\
& \babar~\cite{BaBar:2006jhm}  & 339~\invfb & 1.3$~\times~10^{-7}$ &                            &           &                     \\
& CLEO~\cite{CLEO:1997gxa}     & 4.68~\invfb& 3.7$~\times~10^{-6}$ &                            &           &                     \\\midrule
$\mu^-\pi^0$	       & Belle~\cite{Belle:2007cio}   & 401~\invfb & 1.2$~\times~10^{-7}$ & Belle II~\cite{Belle-II:2022cgf}   & 50~\invab & 7.1$~\times~10^{-10}$ \\
& \babar~\cite{BaBar:2006jhm}  & 339~\invfb & 1.1$~\times~10^{-7}$ &                            &           &                     \\
& CLEO~\cite{CLEO:1997gxa}     & 4.68~\invfb& 4.0$~\times~10^{-6}$ &                            &           &                     \\\midrule
$e^- K_S^0$	       & Belle~\cite{Belle:2010rxj}   & 671~\invfb & 2.6$~\times~10^{-8}$ & Belle II~\cite{Belle-II:2022cgf}   & 50~\invab & 4.0$~\times~10^{-10}$ \\
& \babar~\cite{BaBar:2009qra}  & 469~\invfb & 3.3$~\times~10^{-8}$ &                            &           &                     \\
& CLEO~\cite{CLEO:2002lxe}     & 13.9~\invfb& 9.1$~\times~10^{-7}$ &                            &           &                     \\\midrule

$\mu^- K_S^0$	       & Belle~\cite{Belle:2010rxj}   & 671~\invfb & 2.3$~\times~10^{-8}$ & Belle II~\cite{Belle-II:2022cgf}   & 50~\invab & 4.0$~\times~10^{-10}$ \\
& \babar~\cite{BaBar:2009qra}  & 469~\invfb & 4.0$~\times~10^{-8}$ &                            &           &                     \\
& CLEO~\cite{CLEO:2002lxe}     & 13.9~\invfb& 9.5$~\times~10^{-7}$ &                            &           &                     \\\midrule
$e^- \eta$	       & Belle~\cite{Belle:2007cio}   & 401~\invfb & 9.2$~\times~10^{-8}$ & Belle II~\cite{Belle-II:2022cgf}   & 50~\invab & 1.2$~\times~10^{-9}$  \\
& \babar~\cite{BaBar:2006jhm}  & 339~\invfb & 1.6$~\times~10^{-7}$ &                            &           &                     \\
& CLEO~\cite{CLEO:1997gxa}     & 4.68~\invfb& 8.2$~\times~10^{-6}$ &                            &           &                     \\\midrule

$\mu^-\eta$	       & Belle~\cite{Belle:2007cio}   & 401~\invfb & 6.5$~\times~10^{-8}$ & Belle II~\cite{Belle-II:2022cgf}   & 50~\invab & 8.0$~\times~10^{-10}$ \\
& \babar~\cite{BaBar:2006jhm}  & 339~\invfb & 1.5$~\times~10^{-7}$ &                            &           &                     \\
& CLEO~\cite{CLEO:1997gxa}     & 4.68~\invfb& 9.6$~\times~10^{-6}$ &                            &           &                     \\
\midrule
$e^-\eta^\prime$        & Belle~\cite{Belle:2007cio}   & 401~\invfb & 1.6$~\times~10^{-7}$ & Belle II~\cite{Belle-II:2022cgf}   & 50~\invab & 1.2$~\times~10^{-9}$  \\
& \babar~\cite{BaBar:2006jhm}  & 339~\invfb & 2.4$~\times~10^{-7}$ &                            &           &                     \\\midrule
$\mu^-\eta^\prime$      & Belle~\cite{Belle:2007cio}   & 401~\invfb & 1.3$~\times~10^{-7}$ & Belle II~\cite{Belle-II:2022cgf}   & 50~\invab & 1.2$~\times~10^{-9}$  \\
& \babar~\cite{BaBar:2006jhm}  & 339~\invfb & 1.4$~\times~10^{-7}$ &                            &           &                     \\\midrule
$e^- f_0(980)$	       & Belle~\cite{Belle:2008pdf}   & 671~\invfb & 6.8$~\times~10^{-8}$ & Belle II~\cite{Belle-II:2022cgf}   & 50~\invab & 9.5$~\times~10^{-10}$ \\\hline
$\mu^-f_0(980)$	       & Belle~\cite{Belle:2008pdf}   & 671~\invfb & 6.4$~\times~10^{-8}$ & Belle II~\cite{Belle-II:2022cgf}   & 50~\invab & 9.1$~\times~10^{-10}$ \\\hline
$e^- \rho^0$	       & Belle~\cite{Belle:2011ogy}   & 854~\invfb & 1.8$~\times~10^{-8}$ & Belle II~\cite{Belle-II:2022cgf}   & 50~\invab & 3.8$~\times~10^{-10}$ \\
& \babar~\cite{BaBar:2009wtb}  & 451~\invfb & 4.6$~\times~10^{-8}$ &                            &           &                     \\
& CLEO~\cite{CLEO:1997aqy}     & 4.79~\invfb& 2.0$~\times~10^{-6}$ &                            &           &                     \\

\noalign{\hrule height 1pt}

\end{tabularx}}
\end{adjustwidth}
\end{table}

\begin{table}[H]\ContinuedFloat
\small
\caption{{\em Cont.}}

\begin{adjustwidth}{-\extralength}{0cm}
\setlength{\cellWidtha}{\fulllength/7-2\tabcolsep-0in}
\setlength{\cellWidthb}{\fulllength/7-2\tabcolsep-0in}
\setlength{\cellWidthc}{\fulllength/7-2\tabcolsep-0in}
\setlength{\cellWidthd}{\fulllength/7-2\tabcolsep-0in}
\setlength{\cellWidthe}{\fulllength/7-2\tabcolsep-0in}
\setlength{\cellWidthf}{\fulllength/7-2\tabcolsep-0in}
\setlength{\cellWidthg}{\fulllength/7-2\tabcolsep-0in}
\scalebox{1}[1]{\begin{tabularx}{\fulllength}{>{\PreserveBackslash\centering}m{\cellWidtha}>{\PreserveBackslash\centering}m{\cellWidthb}>{\PreserveBackslash\centering}m{\cellWidthc}>{\PreserveBackslash\centering}m{\cellWidthd}>{\PreserveBackslash\centering}m{\cellWidthe}>{\PreserveBackslash\centering}m{\cellWidthf}>{\PreserveBackslash\centering}m{\cellWidthg}}

\noalign{\hrule height 1pt}

\multicolumn{1}{c}{\textbf{}} & \multicolumn{3}{c}{\textbf{Observed Limits}} & \multicolumn{3}{c}{\textbf{Expected Limits}} \\ \midrule
\boldmath{$\tau^-\to$}            &  \textbf{Experiment}                   & \textbf{Luminosity}       & \textbf{UL (obs)}& \textbf{Experiment} & \textbf{Luminosity}  & \textbf{UL (exp)}        \\ 
\midrule

$\mu^-\rho^0$	       & Belle~\cite{Belle:2011ogy}   & 854~\invfb & 1.2$~\times~10^{-8}$ & Belle II~\cite{Belle-II:2022cgf}   & 50~\invab & 5.5$~\times~10^{-10}$ \\
& \babar~\cite{BaBar:2009wtb}  & 451~\invfb & 2.6$~\times~10^{-8}$ &                            &           &                     \\
& CLEO~\cite{CLEO:1997aqy}     & 4.79~\invfb& 6.3$~\times~10^{-6}$ &                            &           &                     \\\midrule
$e^-\omega$	       & Belle~\cite{Belle:2011ogy}   & 854~\invfb & 4.8$~\times~10^{-8}$ & Belle II~\cite{Belle-II:2022cgf}   & 50~\invab & 1.0$~\times~10^{-9}$  \\
& \babar~\cite{BaBar:2007amy}  & 384~\invfb & 1.1$~\times~10^{-7}$ &                            &           &                     \\       \midrule
$\mu^-\omega$          & Belle~\cite{Belle:2011ogy}   & 854~\invfb & 4.7$~\times~10^{-8}$ & Belle II~\cite{Belle-II:2022cgf}   & 50~\invab & 1.4$~\times~10^{-9}$  \\
& \babar~\cite{BaBar:2007amy}  & 384~\invfb & 1.0$~\times~10^{-7}$ &                            &           &                     \\\midrule
$e^-K^{\ast 0}$	       & Belle~\cite{Belle:2011ogy}   & 854~\invfb & 3.2$~\times~10^{-8}$ & Belle II~\cite{Belle-II:2022cgf}   & 50~\invab & 6.7$~\times~10^{-10}$ \\
& \babar~\cite{BaBar:2009wtb}  & 451~\invfb & 5.9$~\times~10^{-8}$ &                            &           &                     \\
& CLEO~\cite{CLEO:1997aqy}     & 4.79~\invfb& 5.1$~\times~10^{-6}$ &                            &           &                     \\\midrule

$\mu^-K^{\ast 0}$       & Belle~\cite{Belle:2011ogy}   & 854~\invfb & 7.2$~\times~10^{-8}$ & Belle II~\cite{Belle-II:2022cgf}   & 50~\invab & 9.3$~\times~10^{-10}$ \\
& \babar~\cite{BaBar:2009wtb}  & 451~\invfb & 1.7$~\times~10^{-7}$ &                            &           &                     \\
& CLEO~\cite{CLEO:1997aqy}     & 4.79~\invfb& 7.5$~\times~10^{-6}$ &                            &           &                     \\\midrule
$e^-\bar{K}^{\ast0}$    & Belle~\cite{Belle:2011ogy}   & 854~\invfb & 3.4$~\times~10^{-8}$ & Belle II~\cite{Belle-II:2022cgf}   & 50~\invab & 6.2$~\times~10^{-10}$ \\
& \babar~\cite{BaBar:2009wtb}  & 451~\invfb & 4.6$~\times~10^{-8}$ &                            &           &                     \\
& CLEO~\cite{CLEO:1997aqy}     & 4.79~\invfb& 7.4$~\times~10^{-6}$ &                            &           &                     \\\midrule
$\mu^-\bar{K}^{\ast0}$  & Belle~\cite{Belle:2011ogy}   & 854~\invfb & 7.0$~\times~10^{-8}$ & Belle II~\cite{Belle-II:2022cgf}   & 50~\invab & 8.5$~\times~10^{-10}$ \\
& \babar~\cite{BaBar:2009wtb}  & 451~\invfb & 7.3$~\times~10^{-8}$ &                            &           &                     \\
& CLEO~\cite{CLEO:1997aqy}     & 4.79~\invfb& 7.5$~\times~10^{-6}$ &                            &           &                     \\\midrule
$e^-\phi$	       & Belle~\cite{Belle:2011ogy}   & 854~\invfb & 3.1$~\times~10^{-8}$ & Belle II~\cite{Belle-II:2022cgf}   & 50~\invab & 7.4$~\times~10^{-10}$ \\
& \babar~\cite{BaBar:2009wtb}  & 451~\invfb & 3.1$~\times~10^{-8}$ &                            &           &                     \\
& CLEO~\cite{CLEO:1997aqy}     & 4.79~\invfb& 6.9$~\times~10^{-6}$ &                            &           &                     \\\hline
$\mu^-\phi$	       & Belle~\cite{Belle:2011ogy}   & 854~\invfb & 8.4$~\times~10^{-8}$ & Belle II~\cite{Belle-II:2022cgf}   & 50~\invab & 8.4$~\times~10^{-10}$ \\
& \babar~\cite{BaBar:2009wtb}  & 451~\invfb & 1.9$~\times~10^{-7}$ &                            &           &                     \\
& CLEO~\cite{CLEO:1997aqy}     & 4.79~\invfb& 7.0$~\times~10^{-6}$ &                            &           &                     \\\midrule
$e^-e^+e^-$	       & Belle~\cite{Hayasaka:2010np} & 782~\invfb & 2.7$~\times~10^{-8}$ & Belle II~\cite{Belle-II:2022cgf}   & 50~\invab & 4.7$~\times~10^{-10}$ \\
& \babar~\cite{BaBar:2010axs}  & 468~\invfb & 2.9$~\times~10^{-8}$ &                            &           &                     \\
& CLEO~\cite{CLEO:1997aqy}     & 4.79~\invfb& 2.9$~\times~10^{-6}$ &                            &           &                     \\\midrule
$\mu^-e^+e^-$	       & Belle~\cite{Hayasaka:2010np} & 782~\invfb & 1.8$~\times~10^{-8}$ & Belle II~\cite{Belle-II:2022cgf}   & 50~\invab & 2.9$~\times~10^{-10}$ \\
& \babar~\cite{BaBar:2010axs}  & 468~\invfb & 2.2$~\times~10^{-8}$ &                            &           &                     \\
& CLEO~\cite{CLEO:1997aqy}     & 4.79~\invfb& 1.7$~\times~10^{-6}$ &                            &           &                     \\\hline
$e^-\mu^+\mu^-$        & Belle~\cite{Hayasaka:2010np} & 782~\invfb & 2.7$~\times~10^{-8}$ & Belle II~\cite{Belle-II:2022cgf}   & 50~\invab & 4.5$~\times~10^{-10}$  \\
& \babar~\cite{BaBar:2010axs}  & 468~\invfb & 3.2$~\times~10^{-8}$ &                            &           &                     \\
& CLEO~\cite{CLEO:1997aqy}     & 4.79~\invfb& 1.8$~\times~10^{-6}$ &                            &           &                     \\
\midrule
$\mu^-\mu^+\mu^-$      & Belle~\cite{Hayasaka:2010np} & 782~\invfb & 2.1$~\times~10^{-8}$ & Belle II~\cite{Belle-II:2022cgf}   & 50~\invab & 3.6$~\times~10^{-10}$  \\
& \babar~\cite{BaBar:2010axs}  & 468~\invfb & 3.3$~\times~10^{-8}$ &                            &           &                     \\
& LHCb~\cite{LHCb:2014kws}     &   3~\invfb & 4.6$~\times~10^{-8}$ &                            &           &                     \\
& CMS~\cite{CMS:2020kwy}       &  33~\invfb & 8.0$~\times~10^{-8}$ &                            &           &                     \\
& ATLAS~\cite{ATLAS:2016jts}   &  20~\invfb & 3.8$~\times~10^{-7}$ &                            &           &                     \\
& CLEO~\cite{CLEO:1997aqy}     & 4.79~\invfb& 1.9$~\times~10^{-6}$ &                            &           &                     \\\midrule
$e^+\mu^-\mu^-$        & Belle~\cite{Hayasaka:2010np} & 782~\invfb & 1.7$~\times~10^{-8}$ & Belle II~\cite{Belle-II:2022cgf}   & 50~\invab & 2.6$~\times~10^{-10}$  \\
& \babar~\cite{BaBar:2010axs}  & 468~\invfb & 2.6$~\times~10^{-8}$ &                            &           &                     \\
& CLEO~\cite{CLEO:1997aqy}     & 4.79~\invfb& 1.5$~\times~10^{-6}$ &                            &           &                     \\\midrule
$\mu^+e^-e^-$          & Belle~\cite{Hayasaka:2010np} & 782~\invfb & 1.5$~\times~10^{-8}$ & Belle II~\cite{Belle-II:2022cgf}   & 50~\invab & 2.3$~\times~10^{-10}$ \\
& \babar~\cite{BaBar:2010axs}  & 468~\invfb & 1.8$~\times~10^{-8}$ &                            &           &                    \\
& CLEO~\cite{CLEO:1997aqy}     & 4.79~\invfb& 1.5$~\times~10^{-6}$ &                            &           &                    \\

\noalign{\hrule height 1pt}

\end{tabularx}}
\end{adjustwidth}
\end{table}

\begin{table}[H]\ContinuedFloat
\small
\caption{{\em Cont.}}

\begin{adjustwidth}{-\extralength}{0cm}
\setlength{\cellWidtha}{\fulllength/7-2\tabcolsep-0in}
\setlength{\cellWidthb}{\fulllength/7-2\tabcolsep-0in}
\setlength{\cellWidthc}{\fulllength/7-2\tabcolsep-0in}
\setlength{\cellWidthd}{\fulllength/7-2\tabcolsep-0in}
\setlength{\cellWidthe}{\fulllength/7-2\tabcolsep-0in}
\setlength{\cellWidthf}{\fulllength/7-2\tabcolsep-0in}
\setlength{\cellWidthg}{\fulllength/7-2\tabcolsep-0in}
\scalebox{1}[1]{\begin{tabularx}{\fulllength}{>{\PreserveBackslash\centering}m{\cellWidtha}>{\PreserveBackslash\centering}m{\cellWidthb}>{\PreserveBackslash\centering}m{\cellWidthc}>{\PreserveBackslash\centering}m{\cellWidthd}>{\PreserveBackslash\centering}m{\cellWidthe}>{\PreserveBackslash\centering}m{\cellWidthf}>{\PreserveBackslash\centering}m{\cellWidthg}}

\noalign{\hrule height 1pt}

\multicolumn{1}{c}{\textbf{}} & \multicolumn{3}{c}{\textbf{Observed Limits}} & \multicolumn{3}{c}{\textbf{Expected Limits}} \\ \midrule
\boldmath{$\tau^-\to$}            &  \textbf{Experiment}                   & \textbf{Luminosity}       & \textbf{UL (obs)}& \textbf{Experiment} & \textbf{Luminosity}  & \textbf{UL (exp)}        \\ 
\midrule

$e^-\pi^+\pi^-$	       & Belle~\cite{Belle:2012unr}   & 854~\invfb & 2.3$~\times~10^{-8}$ & Belle II~\cite{Belle-II:2022cgf}   & 50~\invab & 5.8$~\times~10^{-10}$ \\
& \babar~\cite{BaBar:2005yvr}  & 221~\invfb & 1.2$~\times~10^{-7}$ &                            &           &                    \\
& CLEO~\cite{CLEO:1997aqy}     & 4.79~\invfb& 2.2$~\times~10^{-6}$ &                            &           &                    \\\midrule
$\mu^-\pi^+\pi^-$      & Belle~\cite{Belle:2012unr}   & 854~\invfb & 2.1$~\times~10^{-8}$ & Belle II~\cite{Belle-II:2022cgf}   & 50~\invab & 5.6$~\times~10^{-10}$ \\
& \babar~\cite{BaBar:2005yvr}  & 221~\invfb & 2.9$~\times~10^{-7}$ &                            &           &                    \\
& CLEO~\cite{CLEO:1997aqy}     & 4.79~\invfb& 8.2$~\times~10^{-6}$ &                            &           &                    \\\midrule
$e^-\pi^+K^-$	       & Belle~\cite{Belle:2012unr}   & 854~\invfb & 3.7$~\times~10^{-8}$ & Belle II~\cite{Belle-II:2022cgf}   & 50~\invab & 7.1$~\times~10^{-10}$ \\
& \babar~\cite{BaBar:2005yvr}  & 221~\invfb & 3.2$~\times~10^{-7}$ &                            &           &                    \\
& CLEO~\cite{CLEO:1997aqy}     & 4.79~\invfb& 6.4$~\times~10^{-6}$ &                            &           &                    \\\midrule
$\mu^-\pi^+K^-$	       & Belle~\cite{Belle:2012unr}   & 854~\invfb & 8.6$~\times~10^{-8}$ & Belle II~\cite{Belle-II:2022cgf}   & 50~\invab & 1.2$~\times~10^{-9}$  \\
& \babar~\cite{BaBar:2005yvr}  & 221~\invfb & 2.6$~\times~10^{-7}$ &                            &           &                    \\
& CLEO~\cite{CLEO:1997aqy}     & 4.79~\invfb& 7.5$~\times~10^{-6}$ &                            &           &                    \\

\midrule
$e^-K^+\pi^{-}$	       & Belle~\cite{Belle:2012unr}   & 854~\invfb & 3.1$~\times~10^{-8}$ & Belle II~\cite{Belle-II:2022cgf}   & 50~\invab & 7.8$~\times~10^{-10}$ \\
& \babar~\cite{BaBar:2005yvr}  & 221~\invfb & 1.7$~\times~10^{-7}$ &                            &           &                    \\
& CLEO~\cite{CLEO:1997aqy}     & 4.79~\invfb& 3.8$~\times~10^{-6}$ &                            &           &                    \\\midrule
$\mu^-K^+\pi^{-}$       & Belle~\cite{Belle:2012unr}   & 854~\invfb & 4.5$~\times~10^{-8}$ & Belle II~\cite{Belle-II:2022cgf}   & 50~\invab & 1.2$~\times~10^{-9}$  \\
& \babar~\cite{BaBar:2005yvr}  & 221~\invfb & 3.2$~\times~10^{-7}$ &                            &           &                    \\
& CLEO~\cite{CLEO:1997aqy}     & 4.79~\invfb& 7.4$~\times~10^{-6}$ &                            &           &                    \\\midrule
$e^-K^+K^-$	       & Belle~\cite{Belle:2012unr}   & 854~\invfb & 3.4$~\times~10^{-8}$ & Belle II~\cite{Belle-II:2022cgf}   & 50~\invab & 6.5$~\times~10^{-10}$ \\
& \babar~\cite{BaBar:2005yvr}  & 221~\invfb & 1.4$~\times~10^{-7}$ &                            &           &                    \\
& CLEO~\cite{CLEO:1997aqy}     & 4.79~\invfb& 6.0$~\times~10^{-6}$ &                            &           &                    \\\midrule
$\mu^-K^+K^-$	       & Belle~\cite{Belle:2012unr}   & 854~\invfb & 4.4$~\times~10^{-8}$ & Belle II~\cite{Belle-II:2022cgf}   & 50~\invab & 1.1$~\times~10^{-9}$  \\
& \babar~\cite{BaBar:2005yvr}  & 221~\invfb & 2.5$~\times~10^{-7}$ &                            &           &                    \\
& CLEO~\cite{CLEO:1997aqy}     & 4.79~\invfb& 1.5$~\times~10^{-5}$ &                            &           &                    \\\midrule
$e^-K_S^0K_S^0$	       & Belle~\cite{Belle:2010rxj}   & 671~\invfb & 7.1$~\times~10^{-8}$ & Belle II~\cite{Belle-II:2022cgf}   & 50~\invab & 9.7$~\times~10^{-10}$ \\
& CLEO~\cite{CLEO:2002lxe}     & 13.9~\invfb& 2.2$~\times~10^{-6}$ &                            &           &                     \\\midrule
$\mu^-K_S^0K_S^0$       & Belle~\cite{Belle:2010rxj}   & 671~\invfb & 8.0$~\times~10^{-8}$ & Belle II~\cite{Belle-II:2022cgf}   & 50~\invab & 1.1$~\times~10^{-9}$  \\
& CLEO~\cite{CLEO:2002lxe}     & 13.9~\invfb& 3.4$~\times~10^{-6}$ &                            &           &                     \\\midrule

$e^+\pi^-\pi^-$        & Belle~\cite{Belle:2012unr}    & 854~\invfb & 2.0$~\times~10^{-8}$ & Belle II~\cite{Belle-II:2022cgf}   & 50~\invab & 4.6$~\times~10^{-10}$ \\
& \babar~\cite{BaBar:2005yvr}  & 221~\invfb & 2.7$~\times~10^{-7}$ &                            &           &                     \\
& CLEO~\cite{CLEO:1997aqy}     & 4.79~\invfb& 1.9$~\times~10^{-6}$ &                            &           &                     \\\midrule
$\mu^+\pi^-\pi^-$      & Belle~\cite{Belle:2012unr}   & 854~\invfb & 3.9$~\times~10^{-8}$ & Belle II~\cite{Belle-II:2022cgf}   & 50~\invab & 4.5$~\times~10^{-10}$ \\
& \babar~\cite{BaBar:2005yvr}  & 221~\invfb & 7.0$~\times~10^{-8}$ &                            &           &                    \\
& CLEO~\cite{CLEO:1997aqy}     & 4.79~\invfb& 3.4$~\times~10^{-6}$ &                            &           &                    \\
\midrule
$e^+\pi^-K^-$          & Belle~\cite{Belle:2012unr}   & 854~\invfb & 3.2$~\times~10^{-8}$ & Belle II~\cite{Belle-II:2022cgf}   & 50~\invab & 7.7$~\times~10^{-10}$ \\
& \babar~\cite{BaBar:2005yvr}  & 221~\invfb & 1.8$~\times~10^{-7}$ &                            &           &                    \\
& CLEO~\cite{CLEO:1997aqy}     & 4.79~\invfb& 2.1$~\times~10^{-6}$ &                            &           &                    \\\midrule
$\mu^+\pi^-K^-$        & Belle~\cite{Belle:2012unr}    & 854~\invfb & 4.8$~\times~10^{-8}$ & Belle II~\cite{Belle-II:2022cgf}   & 50~\invab & 1.2$~\times~10^{-9}$ \\
& \babar~\cite{BaBar:2005yvr}  & 221~\invfb & 2.2$~\times~10^{-7}$ &                            &           &                    \\
& CLEO~\cite{CLEO:1997aqy}     & 4.79~\invfb& 7.0$~\times~10^{-6}$ &                            &           &                    \\\midrule
$e^+K^-K^-$            & Belle~\cite{Belle:2012unr}    & 854~\invfb & 3.3$~\times~10^{-8}$ & Belle II~\cite{Belle-II:2022cgf}   & 50~\invab & 5.8$~\times~10^{-10}$ \\
& \babar~\cite{BaBar:2005yvr}  & 221~\invfb & 1.5$~\times~10^{-7}$ &                            &           &                   \\
& CLEO~\cite{CLEO:1997aqy}     & 4.79~\invfb& 3.8$~\times~10^{-6}$ &                            &           &                    \\\midrule
$\mu^+K^-K^-$          & Belle~\cite{Belle:2012unr}    & 854~\invfb & 4.7$~\times~10^{-8}$ & Belle II~\cite{Belle-II:2022cgf}   & 50~\invab & 9.7$~\times~10^{-10}$ \\
& \babar~\cite{BaBar:2005yvr}  & 221~\invfb & 4.8$~\times~10^{-7}$ &                            &           &                   \\
& CLEO~\cite{CLEO:1997aqy}     & 4.79~\invfb& 6.0$~\times~10^{-6}$ &                            &           &                    \\
\noalign{\hrule height 1pt}

\end{tabularx}}
\end{adjustwidth}
\end{table}

\begin{table}[H]\ContinuedFloat
\small
\caption{{\em Cont.}}

\begin{adjustwidth}{-\extralength}{0cm}
\setlength{\cellWidtha}{\fulllength/7-2\tabcolsep-0in}
\setlength{\cellWidthb}{\fulllength/7-2\tabcolsep-0in}
\setlength{\cellWidthc}{\fulllength/7-2\tabcolsep-0in}
\setlength{\cellWidthd}{\fulllength/7-2\tabcolsep-0in}
\setlength{\cellWidthe}{\fulllength/7-2\tabcolsep-0in}
\setlength{\cellWidthf}{\fulllength/7-2\tabcolsep-0in}
\setlength{\cellWidthg}{\fulllength/7-2\tabcolsep-0in}
\scalebox{1}[1]{\begin{tabularx}{\fulllength}{>{\PreserveBackslash\centering}m{\cellWidtha}>{\PreserveBackslash\centering}m{\cellWidthb}>{\PreserveBackslash\centering}m{\cellWidthc}>{\PreserveBackslash\centering}m{\cellWidthd}>{\PreserveBackslash\centering}m{\cellWidthe}>{\PreserveBackslash\centering}m{\cellWidthf}>{\PreserveBackslash\centering}m{\cellWidthg}}

\noalign{\hrule height 1pt}

\multicolumn{1}{c}{\textbf{}} & \multicolumn{3}{c}{\textbf{Observed Limits}} & \multicolumn{3}{c}{\textbf{Expected Limits}} \\ \midrule
\boldmath{$\tau^-\to$}            &  \textbf{Experiment}                   & \textbf{Luminosity}       & \textbf{UL (obs)}& \textbf{Experiment} & \textbf{Luminosity}  & \textbf{UL (exp)}        \\ \midrule
$\pi^-\bar{\Lambda}$   & Belle~\cite{Belle:2005exq}   & 154~\invfb & 1.4$~\times~10^{-7}$ & Belle II~\cite{Belle-II:2022cgf}   & 50~\invab & 5.5$~\times~10^{-10}$ \\\hline
$\pi^-\Lambda$	       & Belle~\cite{Belle:2005exq}   & 154~\invfb & 7.2$~\times~10^{-8}$ & Belle II~\cite{Belle-II:2022cgf}   & 50~\invab & 5.4$~\times~10^{-10}$ \\\hline
$\bar{p}^-e^+e^-$      & Belle~\cite{Belle:2020lfn}   & 921~\invfb & 3.0$~\times~10^{-8}$ & Belle II~\cite{Belle-II:2022cgf}   & 50~\invab & 4.0$~\times~10^{-10}$ \\\hline
$\bar{p}^-e^+\mu^-$    & Belle~\cite{Belle:2020lfn}   & 921~\invfb & 2.0$~\times~10^{-8}$ & Belle II~\cite{Belle-II:2022cgf}   & 50~\invab & 4.4$~\times~10^{-10}$ \\\hline
$\bar{p}^-\mu^+e^-$    & Belle~\cite{Belle:2020lfn}   & 921~\invfb & 1.8$~\times~10^{-8}$ & Belle II~\cite{Belle-II:2022cgf}   & 50~\invab & 4.4$~\times~10^{-10}$ \\\hline
$\bar{p}^-\mu^+\mu^-$  & Belle~\cite{Belle:2020lfn}   & 921~\invfb & 1.8$~\times~10^{-8}$ & Belle II~\cite{Belle-II:2022cgf}   & 50~\invab & 7.4$~\times~10^{-10}$ \\
& LHCb~\cite{LHCb:2013fsr}     &   1~\invfb & 3.3$~\times~10^{-7}$ &                            &           &                    \\\midrule
$p^+e^-e^-$            & Belle~\cite{Belle:2020lfn}   & 921~\invfb & 3.0$~\times~10^{-8}$ & Belle II~\cite{Belle-II:2022cgf}   & 50~\invab & 3.6$~\times~10^{-10}$ \\\midrule
$p^+\mu^-\mu^-$        & Belle~\cite{Belle:2020lfn}   & 921~\invfb & 4.0$~\times~10^{-8}$ & Belle II~\cite{Belle-II:2022cgf}   & 50~\invab & 8.3$~\times~10^{-10}$ \\
& LHCb~\cite{LHCb:2013fsr}     &   1~\invfb & 4.4$~\times~10^{-7}$ &                            &           &                    \\

\noalign{\hrule height 1pt}
\end{tabularx}}
\end{adjustwidth}

\end{table}

A beam polarization upgrade of the SuperKEKB $e^-e^+$ collider
can enhance the sensitivity to LFV in $\tau$ decays
at the Belle II experiment to levels
beyond the ones listed in Table~\ref{tab:TauDecays}
and Figure~\ref{fig:TauLFV}.
The proposed upgrade~\cite{Banerjee:2022kfy}
will result in ${\sim}70\%$ longitudinal polarization
of the high energy electron beam,
which will influence the angular distribution of the $\tau$ decay products
in the SM $\tau$-pair backgrounds.
The characteristic $\tau$ polarization dependence of
the helicity angles of the $\tau$ decay products
with beam polarization can then be used to  further suppress
the background in, for~example, $\taumg$ searches,
where one $\tau$ decays to a muon and a photon,
while the other $\tau$ decays to a pion and a neutrino,
the decay channel most sensitive to the polarization of the $\tau$ lepton.
Similar background suppression can also be obtained with the other decay modes,
which vary in their sensitivity to the $\tau$ polarization.
In general, the~maximal discriminating power is obtained by studying the polar angles
in the center-of-mass frame times the charge of the $\tau$ decay.

The “irreducible background” from $\tau^- \to \mu^- \nu \bar{\nu} \gamma$
decays are studied in  Figure~\ref{fig:tau-lfv-exp}~\cite{Hitlin:2008gf}.
While the distributions of the backgrounds show marked differences
in the case of beam polarization
with respect to the case of no beam polarization,
the signal distribution modeled by {{uniform}} phase--phase does not change with beam polarization.
{{By removing events where the distribution of the irreducible background shows a rising trend near unity,}}
the background can be reduced significantly,
corresponding to a small loss in signal efficiency.
An optimization study has demonstrated that this would result
in approximately a 10\% improvement in the sensitivity to LFV.
Similar analyses are expected to yield comparable gain in sensitivities
for other decay~modes.

\begin{figure}[H]
\includegraphics[scale=1]{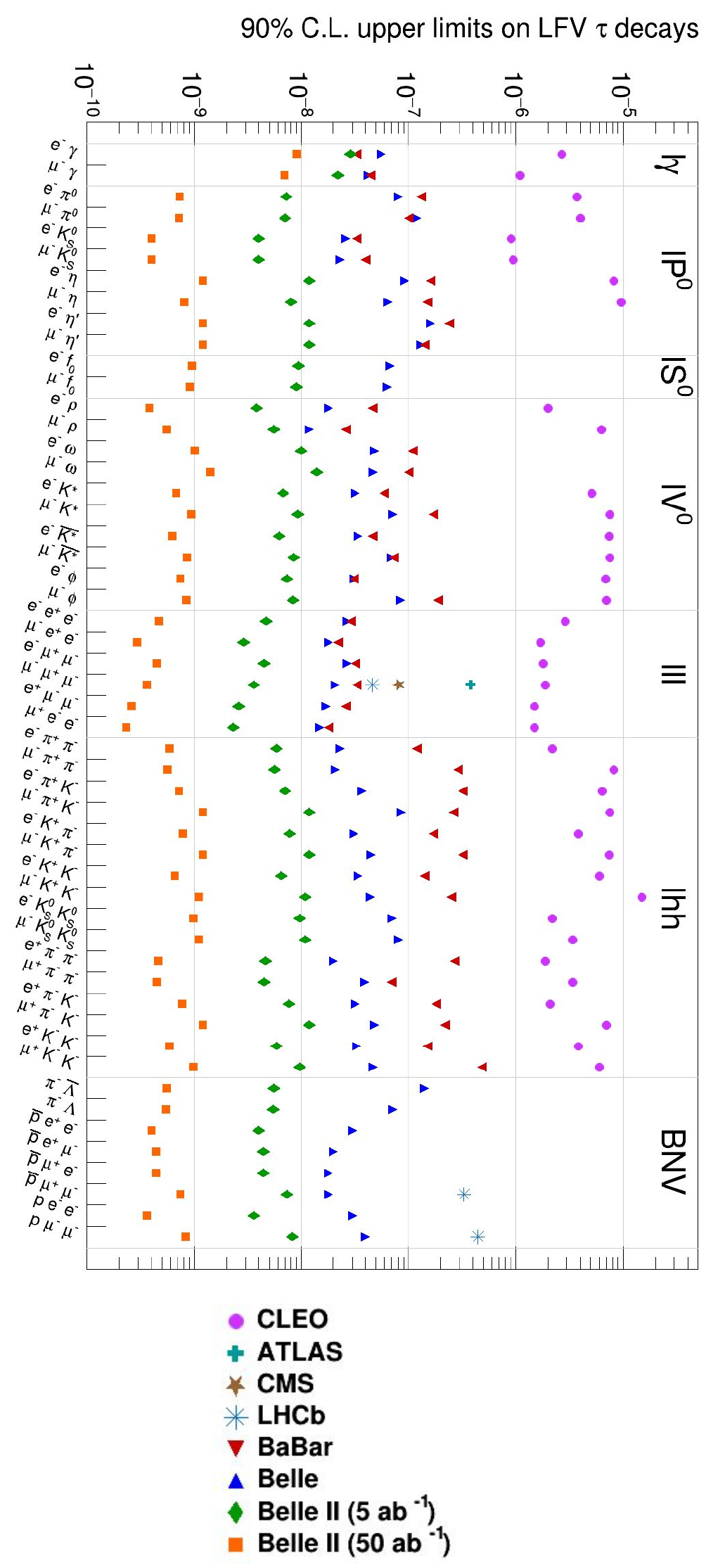}

\caption{
Current status of observed upper limits at CLEO, \babar, Belle, ATLAS, CMS,
and LHCb experiments~\cite{HFLAV:2019otj} 
and projections of expected upper limits
at the Belle II experiment~\cite{Belle-II:2022cgf}.
}
\label{fig:TauLFV}
\end{figure}

\vspace{-6pt}

\begin{figure}[H]
\includegraphics[width=.75\textwidth]{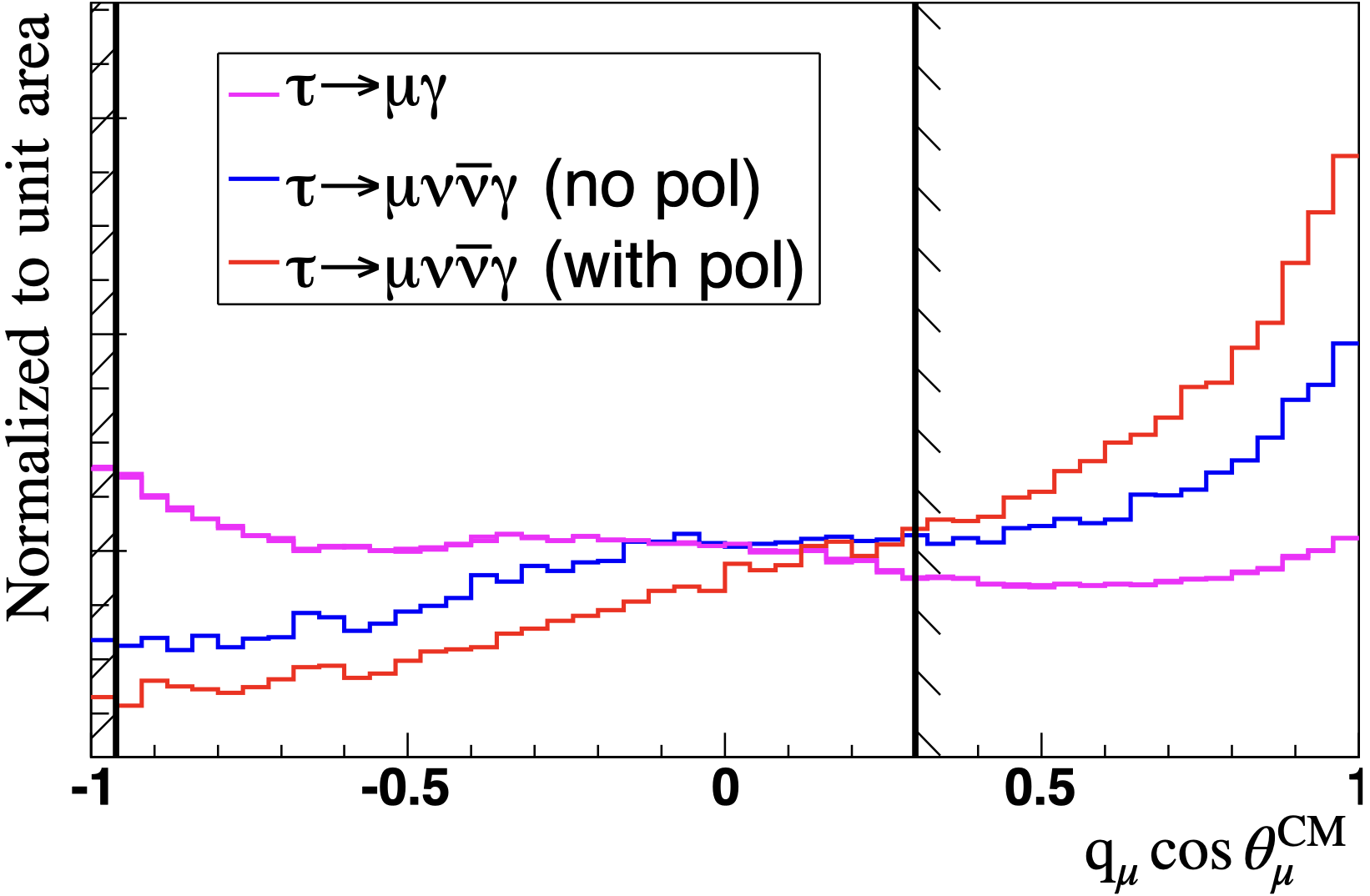} 

\caption{\label{fig:tau-lfv-exp}
Distribution of the cosine of the signal-side muon multiplied by the
muon charge for signal and background events with and without electron
beam polarization in the $\taumg$ search analysis~\cite{Hitlin:2008gf}.
}
\end{figure}

It is worth noting that the uniform phase space model of the signal distribution
is chosen because the underlying theory behind LFV is not known.
Different spin-dependent operators are predicted to give significantly different features
in the Dalitz plane of final state momenta distributions of, for~example,
$\tau^-\to\mu^-\mu^+\mu^-$ decays~\cite{Matsuzaki:2007hh, Dassinger:2007ru}.
One of the most interesting aspects of having the beam polarization
is the possibility to distinguish between these different new physics models
to understand the helicity structure of the couplings producing LFV in $\tau$ decays,
once such decays are~observed.

Belle II limits will probe predictions from several theoretical models,
as listed in Table~\ref{taulfv_predictions}.
Some theoretical expectations from different new physics models and
improvements on experimental limits over the last few decades for
$\taumg$~\cite{Hayes:1981bn,ARGUS:1992ggs,CLEO:1992orj,DELPHI:1995mws,CLEO:1996sqd,CLEO:1999lvl,
Belle:2003ynp,BaBar:2005wms,Belle:2007qih,BaBar:2009hkt,Belle:2021ysv}
and
$\taummm$~\cite{Hayes:1981bn,CLEO:1989ikc,ARGUS:1992ggs,CLEO:1994bbg,CLEO:1997aqy,
Belle:2004vjj,BaBar:2003pme,BaBar:2007yte,Belle:2007diw,BaBar:2010axs,Hayasaka:2010np,
LHCb:2013fsr,LHCb:2014kws,ATLAS:2016jts,CMS:2020kwy} decays,
along with future prospects at Belle II~\cite{Belle-II:2022cgf}, are shown in Figure~\ref{fig:exp_vs_theory}.

\begin{figure}[H]
\centering
\includegraphics[width=.49\textwidth]{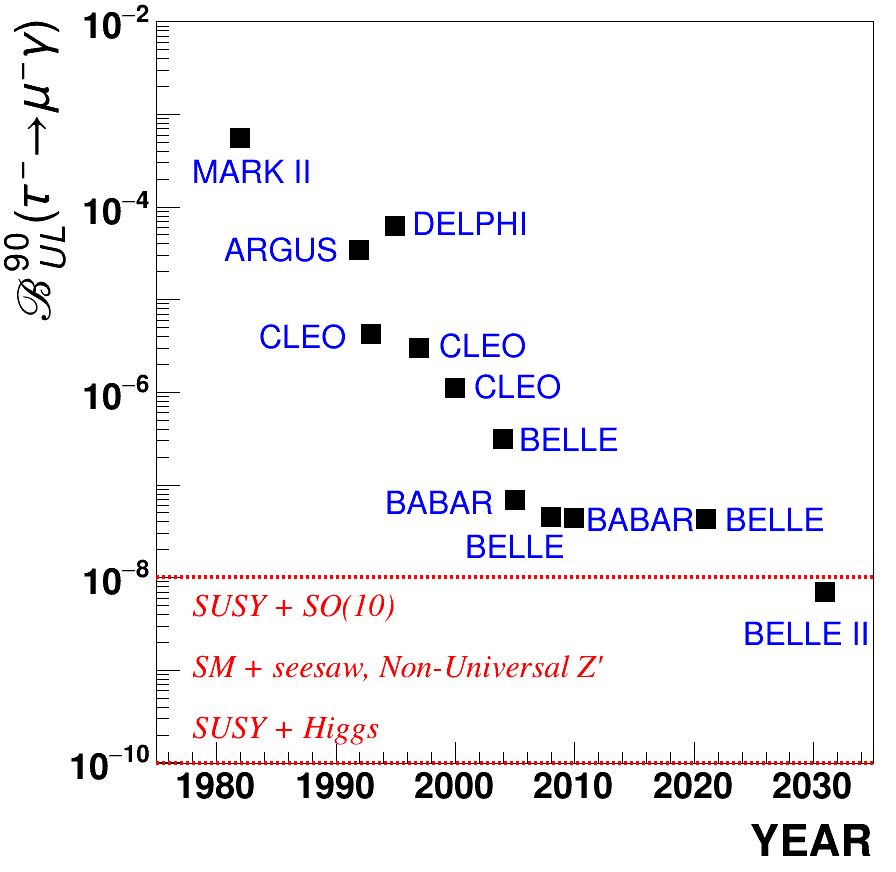}
\includegraphics[width=.49\textwidth]{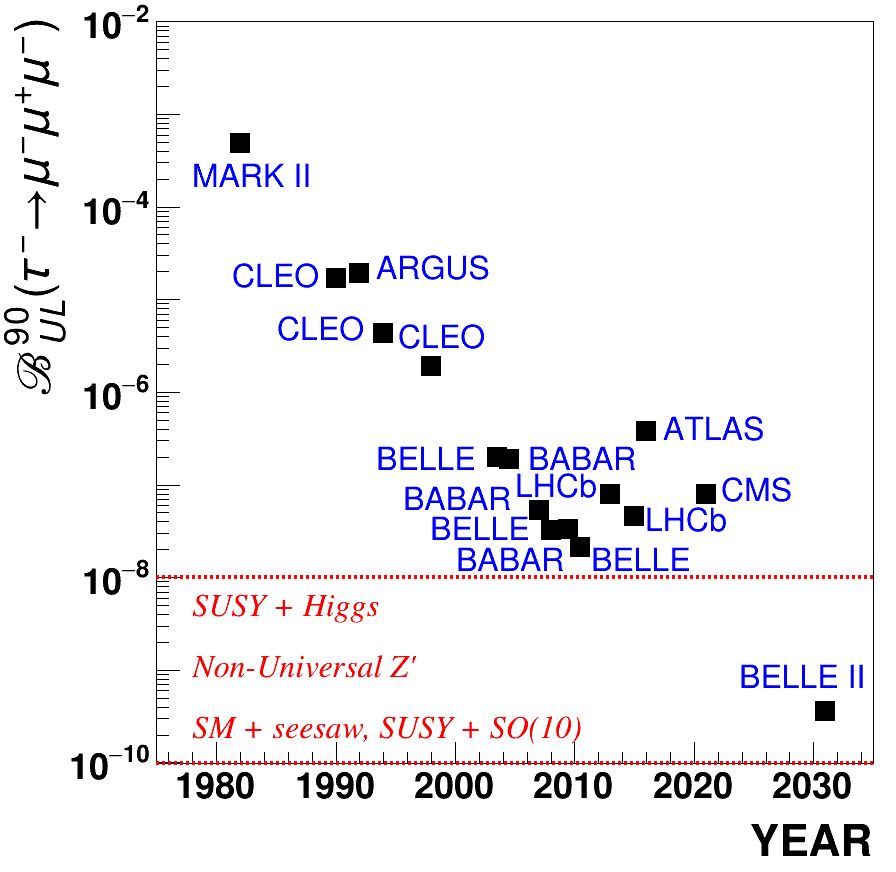}
\caption{Evolution of experimental
bounds on $\taumg$~\cite{Hayes:1981bn,ARGUS:1992ggs,CLEO:1992orj,DELPHI:1995mws,CLEO:1996sqd,CLEO:1999lvl, Belle:2003ynp,BaBar:2005wms,Belle:2007qih,BaBar:2009hkt,Belle:2021ysv} (left)
and $\taummm$~\cite{Hayes:1981bn,CLEO:1989ikc,ARGUS:1992ggs,CLEO:1994bbg,CLEO:1997aqy,
Belle:2004vjj,BaBar:2003pme,BaBar:2007yte,Belle:2007diw,BaBar:2010axs,Hayasaka:2010np,
LHCb:2013fsr,LHCb:2014kws,ATLAS:2016jts,CMS:2020kwy}
decays, future prospects at Belle II~\cite{Belle-II:2022cgf}
and some predictions from Table~\ref{taulfv_predictions}.
}
\label{fig:exp_vs_theory}
\end{figure}
\unskip


\section{Conclusions}

LFV in $\tau$ decays are unambiguous signatures of new physics,
and are thus of great experimental and theoretical interest.
Many models from supersymmetric scenarios to leptoquarks
predict LFV in $\tau$ decays at experimentally observable rates,
which will be probed at Belle II.
Searches for LFV in $\tau$ decays can discover new physics at the multi-TeV scale
by identifying the underlying mechanism beyond the SM,
or strongly constrain the flavor structure of TeV-scale extensions
beyond the SM~\cite{Husek:2020fru,Cirigliano:2021img},
as discussed in the context of different experimental efforts in Ref.~\cite{Banerjee:2022xuw}.
The first generation B-Factory experiments,  Belle and \babar,
saw an order of magnitude improvement on the upper limit on LFV in $\tau$ decays
from $10^{-6}$ level down to $10^{-8}$ level.
The Belle II experiment will continue to improve the sensitivity
in searches of LFV in $\tau$ decays over the next decade.
The projected sensitivity at Belle II for LFV in $\tau$ decays with 50\invab of data
is at the $10^{-10}$--$10^{-9}$ level,
which constitutes one or two orders of magnitude of
improvement over the previous~experiments.

\vspace{6pt}

\funding{This project was supported by the U.S. Department of Energy under research Grant No.~DE-SC0022350.}

\institutionalreview{Not applicable.}

\informedconsent{Not applicable.}

\dataavailability{Not applicable.}

\acknowledgments{
The author acknowledges fruitful discussions with
Kiyoshi Hayasaka,
Kenji Inami,
John Micheal Roney,
Armine Rostomyan,
Michel Hern\'andez Villanueva,
and many others.
Sources of all figures and tables included for discussions have been appropriately cited,
and reproduced with permissions from the respective collaborations.
}

\conflictsofinterest{The author declares no conflict of~interest.}


%

\begin{adjustwidth}{-\extralength}{0cm}
\printendnotes[custom] 

\reftitle{References}



\end{adjustwidth}
\end{document}